\documentclass[twocolumn]{aastex631}

\begin{document}
\title{The Direct-Method Oxygen Abundance of Typical Dwarf Galaxies at Cosmic High-Noon\footnote{The data presented herein were obtained at the W. M. Keck Observatory, which is operated as a scientific partnership among the California Institute of Technology, the University of California, and the National Aeronautics and Space Administration. The Observatory was made possible by the generous financial support of the W. M. Keck Foundation. \\
Based on observations made with the NASA/ESA \textit{Hubble Space Telescope}, obtained from the Data Archive at the Space Telescope Science Institute, which is operated by the Association of Universities for Research in Astronomy, Inc., under NASA contract NAS5-26555. These observations are associated with programs \#9289, \#11710, \#11802, \#12201, \#12931, \#13389, \#14209.}}

\correspondingauthor{Timothy Gburek}
\email{timothy.gburek@email.ucr.edu}

\author[0000-0002-7732-9205]{Timothy Gburek}
\affiliation{Department of Physics \& Astronomy, University of California, Riverside, CA 92521, USA}

\author[0000-0002-4935-9511]{Brian Siana}
\affiliation{Department of Physics \& Astronomy, University of California, Riverside, CA 92521, USA}

\author[0000-0002-8630-6435]{Anahita Alavi}
\affiliation{IPAC, California Institute of Technology, 1200 E. California Boulevard, Pasadena, CA 91125, USA}

\author[0000-0003-2047-1689]{Najmeh Emami}
\affiliation{Minnesota Institute for Astrophysics, University of Minnesota, Minneapolis, MN, 55455, USA}

\author[0000-0001-5492-1049]{Johan Richard}
\affiliation{Univ Lyon, Univ Lyon1, Ens de Lyon, CNRS, Centre de Recherche Astrophysique de Lyon UMR5574, F-69230, Saint-Genis-Laval, France}

\author[0000-0003-3559-5270]{William R. Freeman}
\affiliation{Department of Physics \& Astronomy, University of California, Riverside, CA 92521, USA}

\author{Daniel P. Stark}
\affiliation{Steward Observatory, University of Arizona, 933 N Cherry Ave, Tucson, AZ 85721, USA}

\author[0000-0002-9593-0053]{Christopher Snapp-Kolas}
\affiliation{Department of Physics \& Astronomy, University of California, Riverside, CA 92521, USA}

\begin{abstract}
    \noindent We present a Keck/MOSFIRE, rest-optical, composite spectrum of 16 typical, gravitationally-lensed, star-forming, dwarf galaxies at $1.7 \lesssim z \lesssim 2.6$ ($z_{\rm{mean}}=2.30$), all chosen independent of emission-line strength. These galaxies have a median stellar mass of $\log(M_\ast/\rm{M_\odot})_{\rm{med}} = 8.29^{+0.51}_{-0.43}$ and a median star formation rate of $\rm{SFR_{H\alpha}^{med} = 2.25^{+2.15}_{-1.26}\ M_\odot\ yr^{-1}}$. We measure the faint, electron-temperature-sensitive, [\ion{O}{3}] $\lambda$4363 emission line at $2.5\sigma$ ($4.1\sigma$) significance when considering a bootstrapped (statistical-only) uncertainty spectrum. This yields a direct-method oxygen abundance of $12+\log(\rm{O/H})_{\rm{direct}} = 7.88^{+0.25}_{-0.22}$ ($0.15^{+0.12}_{-0.06}\ \rm{Z_\odot}$). We investigate the applicability at high-$z$ of locally-calibrated, oxygen-based, strong-line metallicity relations, finding that the local reference calibrations of \citet{Bian2018} best reproduce ($\lesssim 0.12$ dex) our composite metallicity at fixed strong-line ratio. At fixed $M_\ast$, our composite is well-represented by the $z \sim 2.3$ direct-method stellar mass$\,-\,$gas-phase metallicity relation (MZR) of \citet{Sanders2020}. When comparing to predicted MZRs from the IllustrisTNG and FIRE simulations, having recalculated our stellar masses with more realistic non-parametric star formation histories (log($M_\ast$/$\rm{M_\odot}$)$_{\rm{med}} = 8.92^{+0.31}_{-0.22}$), we find excellent agreement with the FIRE MZR. Our composite is consistent with no metallicity evolution, at fixed $M_\ast$ and SFR, of the locally-defined fundamental metallicity relation. We measure the doublet ratio [\ion{O}{2}] $\lambda$3729/[\ion{O}{2}] $\lambda3726 = 1.56 \pm 0.32$ ($1.51 \pm 0.12$) and a corresponding electron density of $n_e = 1^{+215}_{-0}\ \rm{cm^{-3}}$ ($n_e = 1^{+74}_{-0}\ \rm{cm^{-3}}$) when considering the bootstrapped (statistical-only) error spectrum. This result suggests that lower-mass galaxies have lower densities than higher-mass galaxies at $z \sim 2$.
\end{abstract}


\section{Introduction} \label{sec:intro}
The gas-phase metallicity, or gas-phase oxygen abundance ($12+\rm{log(O/H)}$) of the interstellar medium (ISM) of galaxies, is a cornerstone in the study of galaxy formation and evolution. The metallicity traces the stellar mass buildup of galaxies through the enrichment over time of the ISM by heavy elements produced via stellar nucleosynthesis. Galaxies, however, are not closed boxes and have inflows of metal-poor gas$\,-\,$the fuel for star formation$\,-\,$ from the circumgalactic medium (CGM) and intergalactic medium (IGM) as well as outflows of metal-laden gas triggered by feedback from supernovae and/or active galactic nuclei (AGN). This modulation of the enrichment of the ISM, via gas flows, shows metallicity to also be an important physical property in the study of the cycle of baryons into, out of, and within (baryon recycling) galaxies.

The combination of these processes is reflected in the scaling relation of the gas-phase metallicity ($Z$) with the stellar mass ($M_\ast$) of star-forming galaxies, known more succinctly as the mass$-$metallicity relation or MZR. This relation demonstrates that stellar mass and metallicity are positively and tightly correlated, whereby in the local Universe, below a characteristic mass of $M_\ast \approx 10^{10.0 - 10.5}\ \rm{M_\odot}$, the MZR is described by a power-law. Above this characteristic mass, or ``turnover" mass, the MZR flattens and asymptotically approaches an upper-limit oxygen abundance. Locally, the MZR has been shown to exist over five decades in $M_\ast$ from $10^6 \lesssim M_\ast/\rm{M_\odot} \lesssim 10^{11}$ \citep[e.g.,][]{Tremonti2004, Lee2006, Kewley&Ellison2008, Andrews&Martini2013, Maiolino&Mannucci2019, Curti2020, Sanders2021}. The MZR has also been shown in numerous studies to exist at high-$z$ out to $z > 3$, though with an evolution such that galaxies at fixed $M_\ast$ have lower metallicities at higher redshifts \citep[e.g.,][]{Erb2006, Maiolino2008, Henry2013_june, Henry2013_oct, Zahid2013, Zahid2014_aug, Zahid2014_sept, Steidel2014, Maiolino&Mannucci2019, Sanders2015, Sanders2020, Sanders2021, Strom2022}. Additionally, at higher redshifts, the turnover mass is found to be larger than seen locally \citep{Zahid2013, Zahid2014_aug, Zahid2014_sept}, and at $z > 2$, it is unknown whether the turnover mass exists at all. At $z > 2$ and $M_\ast \gtrsim 10^9\ \rm{M_\odot}$, the MZR has been described by a single power-law \citep[e.g.,][]{Sanders2021}. 

In constraining the shape, scatter, and evolution in the MZR, insight is gained into the physics of how star formation processes and baryon flows are connected and how galaxy growth is structured and regulated \citep[e.g.,][]{Finlator2008, Dave2012, Ma2016, Torrey2014, Torrey2019}. For instance, the slope of the low-mass end of the MZR can relate galactic metal retention to how efficient outflows (which remove gas and metals from the ISM/galaxy) and stellar feedback are in regulating star formation and stellar mass growth \citep[e.g.,][]{Torrey2014}. Correlated scatter in the MZR can inform of secondary dependencies of the metallicity to properties such as gas-mass and SFR, giving further insight into current conditions of a galaxy as well as elucidating more fundamental relationships between mass, metallicity, and other properties \citep[e.g.,][]{Ma2016, Torrey2019}.

Through the empirical study of the scatter in the MZR, \citet{Mannucci2010} and \citet{Lara-Lopez2010} found that metallicities of galaxies in the Sloan Digital Sky Survey (SDSS) do in fact have a secondary dependence on SFR, a dependence reaffirmed in more recent work, albeit to varying degrees of the strength of that dependence \citep[e.g.,][]{Yates2012, Andrews&Martini2013, Maiolino&Mannucci2019, Curti2020, Sanders2021}. This $M_\ast-\rm{SFR}-\rm{O/H}$ relation is referred to as the fundamental metallicity relation (FMR) and displays a reduced scatter in metallicity of $\sim 0.05$ dex \citep[compared to $\sim 0.1$ dex in the MZR;][]{Tremonti2004}. In effect, the FMR is a 3D surface that posits that metallicity is anti-correlated with SFR such that, at fixed $M_\ast$, galaxies with above-average (below-average) SFRs will have below-average (above-average) O/H. In addition to a reduction in the intrinsic scatter of metallicity, \citet{Mannucci2010} also suggested that the FMR is redshift-invariant out to $z \sim 2.5$. If true, this naturally explains the evolution in the normalization of the MZR to be observations at various redshifts of different regions of the locally-defined FMR; this is physically motivated by the anti-correlation of O/H with SFR and the observed increase of SFR with redshift at fixed $M_\ast$ \citep[e.g.,][]{Speagle2014, Whitaker2014, Sanders2021}. As galaxy samples at high-$z$ have increased in size, evidence has grown that the FMR is indeed redshift-invariant, at least to within $\sim 0.1$ dex, out to $z \sim 2.5$ and even possibly $z \sim 3.3$ \citep[e.g.,][]{Henry2013_june, Henry2013_oct, Cresci2019, Maiolino&Mannucci2019, Sanders2018, Sanders2020, Sanders2021}. However, this evidence is largely based on metallicities indirectly-calculated via prescriptions calibrated in the local Universe, and it is unknown how accurate these methods are at high-$z$.

In order to properly assess the evolution of the MZR and FMR, metallicities must be estimated accurately at low- and high-$z$. This requires an accurate understanding of the nebular physical conditions of star-forming galaxies at different redshifts. Fortunately, a procedure that addresses both of these requirements exists and is applicable at various redshifts, this procedure being the ``direct" method of oxygen abundance determination. This method relies on first estimating the electron temperature ($T_e$) and electron density ($n_e$) of the ionized nebular gas as these properties are responsible for the strength of the collisionally-excited oxygen emission lines needed for this procedure ([\ion{O}{2}] $\lambda\lambda$3726, 3729 and [\ion{O}{3}] $\lambda\lambda$4959, 5007). These properties are then considered together with flux ratios of the collisionally-excited lines to hydrogen Balmer recombination lines in order to estimate the total oxygen abundance \citep[e.g.,][]{Izotov2006, Osterbrock_Ferland2006}. Unfortunately, this direct method relies on weak auroral emission lines to calculate $T_e$, which is determined from the flux ratio of strong emission lines (e.g., [\ion{O}{3}] $\lambda\lambda$4959, 5007) to auroral emission lines (e.g., \ion{O}{3}] $\lambda\lambda$1661, 1666 or [\ion{O}{3}] $\lambda$4363) of the same ionic species. While use of [\ion{O}{3}] $\lambda$4363 is common in this methodology as it lies in the rest-optical with [\ion{O}{3}] $\lambda\lambda$4959, 5007, this line is $\sim30-100\times$ fainter than [\ion{O}{3}] $\lambda$5007 \citep[e.g.,][Figure 1]{Jones2015}, typically decreasing in strength with increasing galactic metallicity. As such, large, representative samples of [\ion{O}{3}] $\lambda$4363-emitters (and thus direct metallicities) have been difficult to acquire with current facilities and instrumentation, especially at high O/H (and $M_\ast$ by the MZR) and at $z > 1$ where only a handful of [\ion{O}{3}] $\lambda$4363 detections exist, mostly thanks to gravitational lensing \citep{Brammer2012, Christensen2012, Stark2013, James2014, Patricio2018, Gburek2019, Sanders2016o3, Sanders2020}. Moreover, due to the faintness of the auroral lines, both in the UV and optical, the currently-detected auroral-line-emitters at $z > 1$ are clearly biased and are more representative of extreme emission line galaxies (EELGs) than of ``typical" star-forming galaxies seen at these redshifts \citep{Sanders2020}. These high-$z$ auroral-line-emitters tend to fall well above the mean $M_\ast-SFR$ relation defined by typical galaxies at a given redshift and have flux ratios indicative of higher ionization parameters and lower metallicities than average.

To overcome the current limitations of the direct-metallicity method, and therein study more representative samples of galaxies across a wider dynamic range of metallicities and redshifts, indirect ``strong-line" methods of determining oxygen abundance were developed \citep{Jensen1976, Alloin1979, Pagel1979}. These methods allow for metallicity estimation when [\ion{O}{3}] $\lambda$4363 cannot be detected. Instead, strong-line methods rely on locally-calibrated empirically \citep[e.g.,][]{Pettini&Pagel2004, Jones2015, Bian2018, Curti2020} or theoretically-determined \citep[e.g.,][]{McGaugh1991, Kewley&Dopita2002, Dopita2013} relations between metallicity and flux ratios of strong, rest-optical, nebular emission lines. However, while these strong-line methods have proven very useful in understanding the enrichment of local galaxies, they have several drawbacks of their own. For example, depending on the strong-line index and calibration used, metallicity estimates can vary by up to 0.7 dex \citep{Kewley&Ellison2008}. In part, this is due to how the strong-line methods are calibrated. Calibrations based on photoionization models tend to produce higher metallicity estimates than empirical, $T_e$-based calibrations \citep[][Figure 3]{Curti2020}. Empirical calibrations can also suffer from sample selection effects whereby individually-detected [\ion{O}{3}] $\lambda$4363-emitters yield metallicities of more extreme star-forming regions whereas metallicities from galaxy samples stacked in order to detect [\ion{O}{3}] $\lambda$4363 may be more representative of ``typical" galaxies \citep{Curti2017, Sanders2020} that fall on the $M_\ast-\rm{SFR}$ relation. 

When considering high-$z$ galaxies, it is unknown if these locally-calibrated strong-line relations, reflective of \ion{H}{2}-region conditions in the local Universe, are applicable for estimating metallicity. Excitation diagrams have shown that star-forming region conditions likely evolve with redshift; this is most notably seen in the [\ion{O}{3}] $\lambda$5007/H$\beta$ vs. [\ion{N}{2}] $\lambda$6583/H$\alpha$ Baldwin$\,-\,$Phillips$\,-\,$Terlevich \citep[N2-BPT;][]{BPT1981} diagram, where the locus of star-forming, high-$z$ galaxies is offset from the locus of local, star-forming, SDSS galaxies \citep[e.g.,][]{Steidel2014, Shapley2015, Strom2017, Strom2018, Kashino2017, Kashino2019, Runco2022}. While it is a current matter of debate as to what is driving this evolution in the locus and thus the \ion{H}{2} region physical conditions (see \citet{Kewley2013} for an analysis of several possibilities such as the ionization parameter, electron density, hardness of the ionizing spectrum, and N/O abundance ratio), it is clear that caution must be taken when applying strong-line metallicity methods at high redshift. Calibrations are needed that are derived from objects with analogous physical conditions to typical, star-forming, high-$z$ galaxies.

In this paper, we analyze a $\langle z \rangle = 2.3$ composite spectrum of 16 gravitationally-lensed, typical, star-forming dwarf galaxies selected independent of emission-line strength. In particular, we study the direct-method metallicity from this composite, derived from a detection of the $T_e$-sensitive [\ion{O}{3}] $\lambda$4363 auroral-line. The paper is organized as follows: In Section \ref{sec:data}, we discuss our observations, data reduction, and sample selection. In Section \ref{sec:measurements}, we discuss our spectral-fitting and stacking methodologies, introduce our composite spectrum, and calculate physical properties of our stacking sample and composite. In Section \ref{sec:results_discussion}, we present our analysis and discussion in regard to how representative our sample is of typical, $z \sim 2.3$, star-forming dwarf galaxies, the applicability of locally-calibrated strong-line metallicity diagnostics at high-$z$, the slope and normalization of the $z \sim 2.3$ MZR, and the redshift evolution of the FMR. In Section \ref{sec:summary_ch3}, we summarize our results. Finally, in Appendix \ref{sec:b18_appendix}, we briefly describe the reasoning and methods behind our refitting of the \citet{Bian2018} strong-line metallicity relations. Throughout this paper, uncertainties reflect our bootstrapped error spectrum for the composite unless stated otherwise. We assume a $\Lambda$CDM cosmology with $H_0$ = 70 km $\rm s^{-1}$ $\rm Mpc^{-1}$, $\Omega_\Lambda$ = 0.7, and $\Omega_m$ = 0.3.

\section{Observations, Data Reduction, and Sample Selection} \label{sec:data}
The focus of this paper is the careful analysis of a stack$\,-\,$from which the oxygen abundance is directly measured$\,-\,$of 16 gravitationally-lensed, star-forming, dwarf galaxies at the peak of cosmic star formation. These galaxies at $1.7 < z < 2.6$ have stellar masses of $\log(M_\ast / M_\odot) < 9.0$ and probe typical dwarf galaxies in this epoch, complimenting the recent large statistical studies of more massive galaxies at these redshifts, such as the Keck Baryonic Structure Survey \citep[KBSS-MOSFIRE;][]{Steidel2014} and the MOSFIRE Deep Evolution Field survey \citep[MOSDEF;][]{Kriek2015}. In this section, we detail the photometric and spectroscopic observations and data reduction of these galaxies and the larger parent surveys from which the galaxies are drawn. We also discuss the selection strategy of these 16 objects chosen for stacking. 

\subsection{Photometric Data and Reduction} \label{subsec:photodata}
The galaxy stacking sample is drawn from a spectroscopic follow-up survey of the photometric \textit{Hubble Space Telescope} (\textit{HST}) survey of \citet{Alavi2014,Alavi2016}, which was conducted to study faint, low-mass, star-forming galaxies gravitationally-lensed by the foreground galaxy clusters Abell 1689, MACS J0717.5+3745, and MACS J1149.5+2223, among others (hereafter A1689, MACS J0717, and MACS J1149, respectively). This \textit{HST} survey compliments the \textit{Hubble} Frontier Fields \citep[HFF;][]{Lotz2017} survey of lensing clusters by both adding deep near-ultraviolet (UV) images of the HFF clusters (of which MACS J0717 and MACS J1149 are members) to the deep HFF optical and near-infrared (IR) datasets as well as by adding or including deep near-UV to near-IR photometry of another lensing cluster, A1689.  

For galaxies lensed by A1689, near-UV images were taken over two programs in the F225W, F275W, and F336W bandpasses with the Wide Field Camera 3 (WFC3)/UVIS channel on the \textit{HST}. As part of Program ID 12201 (PI: B. Siana), F275W was observed for 30 orbits, and F336W was observed for 4 orbits. As part of Program ID 12931 (PI: B. Siana), F336W was observed for an additional 14 orbits (18 orbits total), and F225W was observed for 10 orbits. In the optical, we used existing \textit{HST} photometry, taken with the Advanced Camera for Surveys (ACS)/WFC channel, in the F475W, F625W, F775W, and F850LP bandpasses (PID: 9289; PI: H. Ford) as well as in the F814W bandpass (PID: 11710; PI: J. Blakeslee). A summary of the number of orbits for each near-UV and optical filter, as well as the $5\sigma$ depths for a $0\farcs2$ radius aperture, can be found in \citet[][Table 1]{Alavi2016}. In the near-IR, existing images taken over 1-2 orbits with the F125W and F160W filters and the \textit{HST} WFC3/IR channel (PID: 11802; PI: H. Ford) were used. We note that the near-IR footprint for A1689 is smaller than the near-UV and optical footprints, covering 10 of the 13 stacking sample galaxies (see sample selection in Section \ref{subsec:samplesel}) lensed by A1689.

Galaxies behind the lensing clusters MACS J0717 and MACS J1149 were observed with the WFC3/UVIS channel for 8 orbits in both the F275W and F336W bandpasses as part of the \citet{Alavi2016} \textit{HST} survey under Program ID 13389 (PI: B. Siana). In the optical and near-IR, these clusters were observed with \textit{HST} Director's discretionary time as part of the \textit{Hubble} Frontier Fields survey \citep[HFF;][]{Lotz2017}. As with all clusters in this survey (6 clusters total), MACS J0717 and MACS J1149 were observed for 70 orbits with each ACS/WFC and WFC3/IR (140 orbits total). These HFF clusters are observed in the F435W, F606W, and F814W filters with ACS/WFC and the F105W, F125W, F140W, and F160W filters with WFC3/IR (PID: 13498 for MACS J0717; PID: 13504 for MACS J1149; PI: J. Lotz). Like for the optical and near-UV photometry of A1689, the depths and orbits (both from the HFF survey and other projects) for each filter are listed for MACS J0717 in \citet[][Table 1]{Alavi2016}. This information can be found for MACS J1149 via the Mikulski Archive for Space Telescopes (MAST) website for the HFF survey.\footnote{\url{https://archive.stsci.edu/prepds/frontier/macs1149.html}}

The data reduction, calibration, and photometric measurements for MACS J0717 and MACS J1149 are detailed in \citet{Alavi2016}, as is the UV data reduction and calibration for A1689. The reduction and calibration of the optical data from A1689, as well as the photometric measurements for this cluster, are discussed in \citet{Alavi2014}. The near-IR photometry of A1689 was reduced in the same way as the UV and optical data with the exception that a larger pixel scale of $0\farcs08$ was used in the final drizzled images. As described in \citet{Alavi2014,Alavi2016}, our main photometric catalog for A1689 is built on the UV and optical images with a pixel scale of $0\farcs04$. For the areas of A1689 with near-IR coverage, the multi-band photometry (from UV to near-IR) was remeasured on images with larger pixel scales and that are PSF-matched to the F160W data. The estimations of photometric redshifts, which were used to select the spectroscopic follow-up survey sample detailed in Section \ref{subsec:specdata}, are described in \citet{Alavi2016}.

\subsubsection{Lens Models} \label{subsubsec:lensmodels}
When working with objects gravitationally-lensed by foreground galaxy clusters, accurate lens models are imperative for correcting observed photometry and spectroscopy for the lensing magnification. This correction is necessary for the determination of an object's intrinsic properties (e.g., stellar mass, SFR, etc.). \citet{Alavi2016} detail the lens models considered and used for the HFF clusters and A1689, all of which, while constructed with different assumptions and methodologies, are constrained by the location and redshift of known multiply-imaged systems. As stated in \citet{Alavi2016}, for the HFF clusters we use the lens models derived by the Clusters As TelescopeS (CATS) collaboration,\footnote{\url{https://archive.stsci.edu/prepds/frontier/lensmodels/}} specifically the models of \citet{Limousin2016} and \citet{Jauzac2016} for MACS J0717 and MACS J1149, respectively. For A1689, we use the lens model of \citet{Limousin2007}. These parametric models are all derived via mass reconstruction done with the \texttt{LENSTOOL}\footnote{\url{https://projets.lam.fr/projects/lenstool/wiki}} software \citep{Jullo2007}.

\subsection{Spectroscopic Data and Reduction} \label{subsec:specdata}
As a follow-up to the photometric \textit{HST} survey of \citet{Alavi2014,Alavi2016}, a spectroscopic survey was conducted between 2014 January and 2017 March to obtain near-IR (rest-optical) spectroscopy of select galaxies with the Multi-Object Spectrometer For InfraRed Exploration \citep[MOSFIRE;][]{McLean2010,McLean2012} on the 10 m Keck I telescope. Galaxies for this survey were selected to have high magnifications, observed optical magnitudes (F606W or F625W) less than 26.0 (AB), and photometric redshifts in three redshift ranges, $1.37 \leqslant z \leqslant 1.70$, $2.09 \leqslant z \leqslant 2.61$, and $2.95 \leqslant z \leqslant 3.80$, so that the strong, rest-optical, nebular emission lines of the galaxies lie in the near-IR atmospheric transmission windows. Early selection of galaxies lensed by MACS J0717 and MACS J1149 used photometric redshifts from the CLASH survey \citep{Postman2012}. In all, 151 sources were observed across 9 masks. For galaxies that fall into the two lowest redshift ranges, the strong, nebular emission lines targeted are [\ion{O}{2}] $\lambda\lambda$3726, 3729, H$\beta$, [\ion{O}{3}] $\lambda\lambda$4959, 5007, H$\alpha$, and [\ion{N}{2}] $\lambda\lambda$6548, 6583. To this end, observations of galaxies in the lowest redshift range were conducted using the \textit{Y}-, \textit{J}-, and \textit{H}-band filters, whereas the \textit{J}-, \textit{H}-, and \textit{K}-band filters were used for the two highest redshift ranges. We note that while we targeted the strong, nebular emission lines in the highest redshift range as well, H$\alpha$ and the [\ion{N}{2}] doublet were not observed as they fall outside of the \textit{K}-band's wavelength coverage. 

Observations used an ABBA dither pattern with a $2\farcs5$ dither spacing. The individual exposure time for \textit{J}-band and \textit{H}-band data was 120 s and was 180 s for \textit{Y}-band and \textit{K}-band data. In total, across the 9 masks, the \textit{J}-band was observed between 48 m and 112 m, the \textit{H}-band between 56 m and 112 m, and the \textit{K}-band between 60 m and 120 m, for average total exposure times of 81 m, 85 m, and 82 m, respectively. Data in the \textit{Y}-band were taken for one mask in A1689 for a total of 96 m. In each mask, we used $0\farcs7$-wide slits, yielding spectral resolutions of $R=3388,\ 3318,\ 3660,\ \rm{and}\ 3610$ for the \textit{Y}-, \textit{J}-, \textit{H}-, and \textit{K}-bands, respectively.\footnote{\label{footnote:grating}\url{https://www2.keck.hawaii.edu/inst/mosfire/grating.html}} Our typical FWHM seeing for a given mask/filter combination was $0\farcs71$.

The spectroscopic data obtained with MOSFIRE were reduced with the MOSFIRE Data Reduction Pipeline\footnote{\url{https://keck-datareductionpipelines.github.io/MosfireDRP/}} (DRP). This DRP returns a 2D science spectrum and corresponding 2D error spectrum for each slit in a given mask. Each 2D science spectrum is a composite of the multiple spectra taken at each nod position and is flat-fielded, wavelength-calibrated, background-subtracted, and rectified. For \textit{Y}-, \textit{J}-, and \textit{H}-band spectra, wavelength calibration is performed using the night-sky lines, whereas a combination of night-sky lines and a neon arc lamp is used for \textit{K}-band spectra owing to the faintness of the sky lines and the dominance of thermal noise at the red end of the band. Once the 2D spectra were produced, the 1D spectra were extracted using the custom IDL software \texttt{BMEP}\footnote{\url{https://github.com/billfreeman44/bmep}} from \citet{Freeman2019}. This software is based on the optimal weighting and extraction algorithm of \citet{Horne1986}, with a modification allowing the extraction of fractions of pixels. Each spectrum is flux-calibrated with two stars. A standard star of spectral type B9 V to A2 V is first used to apply a wavelength-dependent calibration. It is ensured that this standard star was observed at an air mass similar to that of the mask under consideration. Following this step, an absolute flux calibration is conducted using a star that was included in the corresponding mask. 

\subsection{Sample Selection for Dwarf Galaxy Stack} \label{subsec:samplesel}
The galaxies that comprise the stack mentioned in the opening of this section are drawn from the photometric and spectroscopic surveys detailed above. These galaxies are required to have a robust spectroscopic redshift and spectroscopic coverage of the strong, rest-optical, nebular emission lines: [\ion{O}{2}] $\lambda\lambda$3726, 3729, H$\beta$, [\ion{O}{3}] $\lambda$4959, H$\alpha$, and [\ion{N}{2}] $\lambda\lambda$6548, 6583. Additionally, these galaxies must have spectroscopic coverage of H$\gamma$ and the faint [\ion{O}{3}] $\lambda$4363 auroral emission line. The auroral line is essential for determining gas-phase metallicity directly as it is a component of the emission-line ratio used to estimate electron temperature ($T_e$; see Section \ref{subsec:Te_ne} for more details). These redshift and coverage requirements yield a sample of 18 galaxies and 24 total spectra when accounting for multiply-imaged systems, of which we have four in our sample. A final cut is made on stellar mass (see Section \ref{subsec:SED} on mass estimation) to ensure that our sample lies in the dwarf galaxy regime ($\log(M_\ast / M_\odot) < 9.0$). With this cut, two galaxies are removed from our sample, yielding a final count of 16 galaxies (22 total spectra) ranging in redshift from $z=1.70$ to $z=2.59$ ($z_{\rm mean}=2.30$).

We note here that H$\gamma$ coverage is included as a requirement so as to provide another Balmer decrement with which to estimate the dust extinction from the stack. Due to the close proximity of H$\gamma$ (4340 $\rm{\AA}$) to [\ion{O}{3}] $\lambda$4363, this inclusion does not affect our sample size. We also note here that we do not require spectroscopic coverage of the [\ion{O}{3}] $\lambda$5007 line of the [\ion{O}{3}] $\lambda\lambda$4959, 5007 doublet so as to maximize our galaxy count by including those sources for which [\ion{O}{3}] $\lambda$5007 falls just redward of a given filter. Instead, when necessary, we make use of the $T_e$-insensitive intrinsic intensity ratio of the doublet: [\ion{O}{3}] $\lambda$5007/[\ion{O}{3}] $\lambda$4959 = 2.98 \citep{Storey&Zeippen2000}. Lastly, while we do not select galaxies based on the strength of any given emission line, we do note that each spectrum has a signal-to-noise ratio (S/N) for [\ion{O}{3}] $\lambda$5007 of $\rm{S/N} > 5$, ensuring accurate normalization of each spectrum (by [\ion{O}{3}] $\lambda$5007) during the stacking process (see Section \ref{subsec:stack}). A summary of our sample, and some of the galaxies' physical properties, are listed in Table \ref{tab:stacking_sample}.

\begin{deluxetable*}{clccccccc}[ht!]
\vspace{0.1cm}
\tablecaption{Summary and Properties of $\langle z \rangle = 2.3$ Dwarf Galaxy Stacking Sample\label{tab:stacking_sample}}
\tablecolumns{9}
\tablenum{1}
\tablewidth{\textwidth}
\setlength{\tabcolsep}{6.3pt}
\renewcommand{\arraystretch}{1.3}
\tablehead{
\colhead{Galaxy} &
\colhead{Spec. ID} &
\colhead{$z\tablenotemark{a}$} &
\colhead{R.A.$\tablenotemark{b}$} &
\colhead{Dec.$\tablenotemark{b}$} &
\colhead{$\log(\frac{M_\ast}{\rm{M_\odot}})_{\rm{fid}}\tablenotemark{c}$} &
\colhead{$\log(\frac{M_\ast}{\rm{M_\odot}})\tablenotemark{d}$} &
\colhead{$\frac{\rm{SFR}}{\rm{M_\odot\ yr^{-1}}}\tablenotemark{e}$} &
\colhead{$\log(\frac{L_{\rm{H\alpha}}}{\rm{erg\ s^{-1}}})\tablenotemark{f}$}
}
\startdata
1  & A1689-1037 & 1.70089 & 13:11:35.197 & -01:20:25.040 & $7.71^{+0.20}_{-0.37}$ & $8.12^{+0.22}_{-0.23}$ & $0.151 \pm 0.004$ & $40.511^{+0.013}_{-0.013}$ \\
2  & A1689-1197 & 1.70261 & 13:11:29.689 & -01:20:08.769 & $8.45^{+0.05}_{-0.05}$ & $8.87^{+0.13}_{-0.10}$ & $-\tablenotemark{h}$ & $-\tablenotemark{h}$ \\
   & A1689-370  & 1.70257 & 13:11:32.406 & -01:21:16.027 & $8.31^{+0.05}_{-0.05}$ & $8.71^{+0.12}_{-0.11}$ & $1.779 \pm 0.020$ & $41.583^{+0.005}_{-0.005}$ \\
   & \textit{Composite}$\tablenotemark{g}$  & 1.70259 & $-$ & $-$ & $8.36^{+0.03}_{-0.04}$ & $8.79^{+0.08}_{-0.08}$ & $1.779 \pm 0.020$ & $41.583^{+0.005}_{-0.005}$ \\
3  & A1689-280  & 1.70316 & 13:11:31.886 & -01:21:26.014 & $7.91^{+0.04}_{-0.05}$ & $8.70^{+0.23}_{-0.21}$ & $0.692 \pm 0.011$ & $41.173^{+0.007}_{-0.007}$ \\
4  & A1689-257  & 1.70355 & 13:11:26.426 & -01:21:31.277 & $7.81^{+0.08}_{-0.10}$ & $8.47^{+0.18}_{-0.14}$ & $1.809 \pm 0.027$ & $41.590^{+0.006}_{-0.007}$ \\
5  & A1689-1751 & 2.38159 & 13:11:31.333 & -01:19:18.559 & $8.82^{+0.07}_{-0.09}$ & $9.23^{+0.17}_{-0.12}$ & $3.071 \pm 0.145$ & $41.820^{+0.020}_{-0.021}$ \\
6  & A1689-232  & 2.38709 & 13:11:32.794 & -01:21:27.893 & $7.06^{+1.10}_{-7.06}$ & $8.70^{+0.20}_{-0.19}$ & $2.368 \pm 0.185$ & $41.707^{+0.033}_{-0.035}$ \\
7  & M0717-3958 & 2.39329 & 07:17:27.442 & +37:45:25.475 & $8.69^{+0.11}_{-0.16}$ & $9.42^{+0.09}_{-0.12}$ & $2.072 \pm 0.139$ & $41.649^{+0.028}_{-0.030}$ \\
   & M0717-4517 & 2.39330 & 07:17:27.050 & +37:45:09.695 & $8.99^{+0.07}_{-0.09}$ & $9.46^{+0.09}_{-0.09}$ & $2.350 \pm 0.128$ & $41.704^{+0.023}_{-0.024}$ \\
   & \textit{Composite}$\tablenotemark{g}$  & 2.39329 & $-$ & $-$ & $8.84^{+0.07}_{-0.08}$ & $9.44^{+0.07}_{-0.07}$ & $2.223 \pm 0.094$ & $41.680^{+0.018}_{-0.019}$ \\
8  & A1689-1059 & 2.41141 & 13:11:25.228 & -01:20:19.309 & $8.40^{+0.03}_{-0.03}$ & $9.03^{+0.17}_{-0.21}$ & $5.041 \pm 0.464$ & $42.035^{+0.038}_{-0.042}$ \\
9  & A1689-1467 & 2.51903 & 13:11:26.118 & -01:19:42.837 & $8.21^{+0.10}_{-0.13}$ & $8.76^{+0.18}_{-0.17}$ & $2.287 \pm 0.441$ & $41.692^{+0.077}_{-0.093}$ \\
10 & A1689-1216 & 2.54082 & 13:11:31.981 & -01:20:07.173 & $8.70^{+0.06}_{-0.07}$ & $9.07^{+0.15}_{-0.15}$ & $0.822 \pm 0.144$ & $41.248^{+0.070}_{-0.084}$ \\
   & A1689-1292 & 2.54064 & 13:11:26.528 & -01:19:55.146 & $8.90^{+0.06}_{-0.07}$ & $9.07^{+0.11}_{-0.13}$ & $0.667 \pm 0.097$ & $41.157^{+0.059}_{-0.068}$ \\
   & A1689-537  & 2.54046 & 13:11:29.795 & -01:21:05.969 & $8.60^{+0.08}_{-0.10}$ & $9.05^{+0.17}_{-0.19}$ & $1.725 \pm 0.397$ & $41.570^{+0.090}_{-0.114}$ \\
   & \textit{Composite}$\tablenotemark{g}$  & 2.54065 & $-$ & $-$ & $8.70^{+0.04}_{-0.04}$ & $9.07^{+0.08}_{-0.08}$ & $0.755 \pm 0.079$ & $41.211^{+0.043}_{-0.048}$ \\
11 & A1689-470  & 2.54112 & 13:11:26.213 & -01:21:09.695 & $7.28^{+0.31}_{-7.28}$ & $8.49^{+0.22}_{-0.21}$ & $1.235 \pm 0.160$ & $41.425^{+0.053}_{-0.060}$ \\
12 & A1689-1451 & 2.54201 & 13:11:28.682 & -01:19:42.849 & $8.22^{+0.08}_{-0.09}$ & $9.10^{+0.17}_{-0.16}$ & $0.756 \pm 0.128$ & $41.211^{+0.068}_{-0.080}$ \\
13 & A1689-722  & 2.54247 & 13:11:33.915 & -01:20:52.526 & $8.78^{+0.20}_{-0.36}$ & $8.81^{+0.22}_{-0.24}$ & $3.771 \pm 0.291$ & $41.910^{+0.032}_{-0.035}$ \\
14 & M0717-1531 & 2.55185 & 07:17:32.547 & +37:45:02.348 & $9.08^{+0.11}_{-0.14}$ & $9.94^{+0.25}_{-0.06}$ & $5.558 \pm 0.441$ & $42.078^{+0.033}_{-0.036}$ \\
   & M0717-3187 & 2.55159 & 07:17:35.089 & +37:45:48.120 & $8.90^{+0.10}_{-0.14}$ & $9.67^{+0.17}_{-0.15}$ & $5.965 \pm 0.268$ & $42.109^{+0.019}_{-0.020}$ \\
   & M0717-5970 & 2.55167 & 07:17:30.613 & +37:44:22.798 & $8.82^{+0.17}_{-0.29}$ & $9.66^{+0.08}_{-0.10}$ & $3.969 \pm 0.283$ & $41.932^{+0.030}_{-0.032}$ \\
   & \textit{Composite}$\tablenotemark{g}$ & 2.55168 & $-$ & $-$ & $8.93^{+0.07}_{-0.09}$ & $9.72^{+0.07}_{-0.07}$ & $5.109 \pm 0.178$ & $42.041^{+0.015}_{-0.015}$ \\
15 & A1689-217  & 2.59181 & 13:11:27.623 & -01:21:35.622 & $8.23^{+0.04}_{-0.04}$ & $9.22^{+0.17}_{-0.18}$ & $9.194 \pm 0.313$ & $42.297^{+0.015}_{-0.015}$ \\
16 & M1149-2185 & 2.59366 & 11:49:40.162 & +22:25:07.571 & $8.83^{+0.05}_{-0.06}$ & $9.33^{+0.04}_{-0.04}$ & $33.602 \pm 3.037$ & $42.859^{+0.038}_{-0.041}$
\enddata
\tablenotetext{a}{Spectroscopic redshift$\,-\,$all uncertainties are $\sigma_z \lesssim 6 \times 10^{-5}$.}
\tablenotetext{b}{Right Ascension: hh:mm:ss.sss; Declination: dd:mm:ss.sss; Equinox: J2000}
\tablenotetext{c}{Our fiducial de-magnified stellar mass estimates assuming constant star formation histories (SFH). See Section \ref{subsec:SED}.}
\tablenotetext{d}{De-magnified stellar masses estimated assuming non-parametric SFHs. These estimates are considered in Section \ref{subsubsec:sim_mzrs}.}
\tablenotetext{e}{Star formation rates calculated from H$\alpha$ luminosities assuming a \citet{Chabrier2003} IMF. See Section \ref{subsec:dust_sfr} and Equation \ref{equ:SFR}.}
\tablenotetext{f}{All H$\alpha$ luminosities are corrected for slit-loss (Section \ref{subsubsec:SLC}), magnification (Section \ref{subsubsec:lensmodels}), and dust extinction (Section \ref{subsec:dust_sfr}). The dust extinction correction applied to each luminosity is the same and is derived from the composite spectrum of the total sample. The luminosities are not corrected for stellar absorption, which on average would result in an increase of $< 1\%$.} 
\tablenotetext{g}{The redshift, stellar masses, and H$\alpha$ luminosity are weighted-averages. The SFR is calculated from this luminosity.}
\tablenotetext{h}{H$\alpha$ was not used here or in the composite of A1689-1197 and A1689-370 because it is at the edge of our $H$-band spectrum of this image (with [\ion{N}{2}] $\lambda$6583 falling outside of our coverage).}
\end{deluxetable*}

\section{Measurements and Stacking Methodology} \label{sec:measurements}
In this section, we detail our methodologies for fitting the spectroscopy and photometry of the dwarf galaxies in our stacking sample. We also discuss how various physical properties are estimated either for the individual galaxies or for the ``sample-average" dwarf galaxy, represented by a composite spectrum of these dwarfs. We begin by discussing the measurements made for individual galaxies and then proceed to the construction and analysis of the composite spectrum.

\subsection{Fitting the Individual Emission-Line Spectra} \label{subsec:fit_ind_spectra}
Each emission-line spectrum in our stacking sample (22 total), corresponding to either the single image of a galaxy or one of a multiply-imaged galaxy, is fit using the Markov Chain Monte Carlo (MCMC) Ensemble sampler \texttt{emcee}\footnote{\url{https://emcee.readthedocs.io/en/v2.2.1/}} \citep{Foreman-Mackey2013}. The best-fit model of each spectrum is informed by the science spectrum and corresponding error spectrum and is the model with the maximum likelihood. The general model used in this work is comprised of a line fit to each spectrum's continuum (which are not significantly detected), and single-Gaussian profiles fit to the emission lines. To minimize the impact on the spectral-fitting from pixels contaminated by sky lines, we removed, prior to fitting, any pixels with a corresponding error spectrum value $>3\times$ the median error value over the range of the fit. 

When fitting, each spectroscopic band (\textit{Y}, \textit{J}, \textit{H}, \textit{K}) was considered separately. For each spectrum, the slope and intercept of the continuum were free parameters. In the band containing H$\alpha$ and the [\ion{N}{2}] doublet (the \textit{H}- or the \textit{K}-band), the free parameters also included the redshift of the spectrum, the width of the emission lines (each line having the same width), and the amplitudes of the H$\alpha$ and [\ion{N}{2}] $\lambda$6583 lines, with the amplitude of [\ion{N}{2}] $\lambda$6583 constrained such that [\ion{N}{2}] $\lambda$6583/[\ion{N}{2}] $\lambda$6548 = 2.95 \citep{Acker1989}.

In the band containing H$\gamma$, [\ion{O}{3}] $\lambda$4363, H$\beta$, and [\ion{O}{3}] $\lambda$4959 (the \textit{J}- or \textit{H}-band), two fits were conducted due to the large wavelength separation between [\ion{O}{3}] $\lambda$4363 and H$\beta$. The portion of the spectrum containing H$\beta$ and [\ion{O}{3}] $\lambda$4959 (and [\ion{O}{3}] $\lambda$5007 if covered) was fit first, having the free parameters of line-width for the filter, redshift, and emission-line amplitudes. If [\ion{O}{3}] $\lambda$5007 is within the spectrum's wavelength coverage, its amplitude was fit with the constraint that [\ion{O}{3}] $\lambda$5007/[\ion{O}{3}] $\lambda$4959 = 2.98 \citep{Storey&Zeippen2000}. Otherwise, the amplitude of [\ion{O}{3}] $\lambda$4959 was fit, and the line's flux was multiplied by the aforementioned intensity ratio in order to estimate the [\ion{O}{3}] $\lambda$5007 flux. With a best-fit width and redshift in-hand from the first fit to the filter, the fainter H$\gamma$ and [\ion{O}{3}] $\lambda$4363 lines were then fit with these two parameters fixed. 

Finally, in the band containing the [\ion{O}{2}] $\lambda\lambda$3726, 3729 doublet (the \textit{Y}- or \textit{J}-band), the redshift and line-width were fixed to the values fit to the complete spectrum's highest S/N line in order to avoid complications resulting from the doublet lines' small wavelength separation. In addition to the continuum parameters, only the [\ion{O}{2}] lines' amplitudes were free parameters in these fits. 

Ultimately, the final redshift given to the full spectrum is the weighted-average of the redshifts fit to the \textit{J} (\textit{H})- and \textit{H} (\textit{K})-bands. The flux of a given emission line is found via the equation $f=\sqrt{2\pi}A\sigma$, where $A$ is the emission line's amplitude, and $\sigma$ is its line-width. Since MCMC fitting involves a chain of values for each free parameter, generally $A$ and $\sigma$ in the flux equation, a chain of flux values results for each emission line. The best-fit flux and its uncertainty for each line is then taken to be the most probable value of the line's flux distribution (or posterior) and the posterior's $1\sigma$ width, respectively. 

In regard to the hydrogen Balmer emission lines, the measured line fluxes relative to a linear continuum model are underestimated as they do not account for Balmer absorption in the atmospheres of (primarily A-type) stars. This absorption is present in each spectrum's real stellar continuum and is coincident with the nebular Balmer emission lines. We estimate the H$\gamma$, H$\beta$, and H$\alpha$ absorption corrections in each spectrum with the slit-loss-corrected (see Section \ref{subsubsec:SLC}) line profiles fit to the Balmer emission lines in combination with the model continuum derived for each spectrum with our SED-fitting (see Section \ref{subsec:SED}). For H$\gamma$, H$\beta$, and H$\alpha$, we find sample-median stellar absorption corrections of $\sim5.0\%$, $\sim1.7\%$, and $\sim0.4\%$, respectively, which are used to correct (increase) the Balmer emission-line luminosities of our stacking sample's composite spectrum (see Section \ref{subsec:stack}). It is with these corrected, composite, Balmer emission lines that we estimate extinction due to nebular dust (see Section \ref{subsec:dust_sfr}).

\subsubsection{Slit-Loss Correction} \label{subsubsec:SLC}
When measuring the emission-line fluxes from spectra observed through slit masks, care must be taken to account for loss of flux outside of the slits in order to recover the true integrated flux values. To this end, our line-fluxes were slit-loss-corrected on a galaxy-by-galaxy basis using the methodology of \citet{Emami2020}. 

\subsection{SED-Fitting and Stellar Mass Estimation} \label{subsec:SED}
To determine the stellar masses of the galaxies in our stacking sample, we fit spectral energy distributions (SEDs) to our \textit{HST} near-UV to near-IR photometry (we note that three galaxies lensed by A1689 lack near-IR photometry; see Section \ref{subsec:photodata}). At high redshift, observations suggest that high equivalent width emission lines are fairly common, particularly in lower-mass galaxies like those in our sample \citep{Reddy2018_EW}. Therefore, prior to SED-fitting, we subtracted off any contribution to the photometry from the slit-loss-corrected, nebular emission lines. We also added an additional $3\%$ flux error, in quadrature, to all bands in order to account for systematic errors in the photometry \citep{Alavi2016}. To this emission-line-corrected photometry, we then fit \citet{BC2003} stellar population synthesis models using the SED-fitting code \texttt{FAST}\footnote{\url{https://w.astro.berkeley.edu/~mariska/FAST.html}} \citep{Kriek2009}. We assume constant star formation histories (SFH), a \citet{Chabrier2003} initial mass function (IMF), stellar metallicities of 0.2 $\rm{Z_\odot}$ or 0.4 $\rm{Z_\odot}$, and a \citet{Calzetti2000} dust attenuation curve. The redshifts of the galaxies are fixed to their fit spectroscopic values. We note that our assumption here of constant SFHs is made in order to be generally consistent with the SED-fitting methodologies in relevant literature as this allows more direct comparison of our empirical results. We revisit and revise this assumption in Section \ref{subsubsec:sim_mzrs} when discussing our results against those from cosmological simulations, recalculating our stellar masses assuming less simplistic, more realistic, non-parametric SFHs. Our stellar mass estimates under either SFH assumption are listed in Table \ref{tab:stacking_sample}.

Uncertainties on the properties estimated by \texttt{FAST} (e.g., stellar mass, SFR, $A_V$, etc.) are derived using a Monte Carlo approach where the photometry being fit is perturbed based on its uncertainties and is then refit, this process being repeated 300 times. From these 300 realizations of the SED, $68\%$ confidence intervals are determined for each estimated property. In Table \ref{tab:stacking_sample}, we list the best-fit stellar mass, and its uncertainty, of each galaxy in our stacking sample. The stellar mass associated with our full-sample composite, detailed below, is taken to be the median of these individual masses, log($M_\ast$/$\rm{M_\odot}$)$_{\rm{med}}$ = $8.29^{+0.51}_{-0.43}$, with the stated error bars representing the interquartile range (IQR) of the masses. We note that the best-fit SEDs, stellar masses, and all other affected properties are de-magnified based on the lensing models discussed in Section \ref{subsubsec:lensmodels}.

\subsection{The Composite Spectrum} \label{subsec:stack}
While the individual galaxies in our sample display several nebular emission lines at high-S/N (e.g., [\ion{O}{3}], H$\alpha$), the galaxies are still inherently faint even with high magnification via gravitational-lensing. As a result, many other useful, fainter lines are undetected or marginally-detected in our individual spectra. Such lines can include H$\beta$ and H$\gamma$ for estimating extinction from dust, the [\ion{O}{2}] doublet for calculating electron density, and especially [\ion{O}{3}] $\lambda$4363 for estimating electron temperature and metallicity directly. A composite spectrum, or stacked spectrum, of all of our sample galaxies offers a solution to this problem by including in our study both galaxies for which we have individual line detections and galaxies for which we only have upper limits. This composite gives the advantages of both increasing the S/N of faint spectral features and displaying the average spectrum and properties of dwarf galaxies like those in our sample. Additionally, we use composites of the individual spectra of galaxies multiply-imaged by lensing in order to increase the effective exposure times and S/N of those galaxies' spectroscopy.

Our methodology for creating composite spectra is similar when stacking multiple images of sources (4 multiply-imaged galaxies; see Table \ref{tab:stacking_sample}) or all of the galaxies in our sample (16 total). We first create the composites for our multiply-imaged galaxies as these composites represent their corresponding galaxies in the full-sample stack. For any stack, we begin by shifting the slit-loss-corrected, observed spectra to the rest-frame and converting the flux densities into luminosity densities assuming the corresponding fit spectroscopic redshifts as fixed in either process. Each spectrum is then normalized by its slit-loss-corrected, [\ion{O}{3}] $\lambda$5007 emission-line luminosity. This normalization serves two purposes. It de-magnifies each spectrum implicitly by dividing the magnified spectrum by its magnified [\ion{O}{3}] $\lambda$5007 luminosity. It also, in the case of stacking our full sample, prevents our composite electron temperature (see Section \ref{subsec:Te_ne}) from being biased by the brightest [\ion{O}{3}] $\lambda$5007 source \citep{Sanders2020}. We note here that, prior to normalizing, the spectra and [\ion{O}{3}] $\lambda$5007 luminosities are not corrected for dust extinction both due to the faintness of H$\beta$ and H$\gamma$ and the sky line contamination of these lines in several of our individual sources. (We discuss dust-correcting the full-sample composite as well as the H$\alpha$ luminosities of individual sources when estimating their SFRs in Section \ref{subsec:dust_sfr}.) Following normalization, each spectroscopic band's science spectrum and propagated $1\sigma$ error spectrum are resampled with the Python tool \texttt{SpectRes}\footnote{\url{https://spectres.readthedocs.io/en/latest/}}$^{,}$\footnote{\url{https://github.com/ACCarnall/SpectRes}} \citep{Carnall2017} onto a common wavelength grid with a rest-frame dispersion $-$ for the full-sample stack $-$ of 0.38 $\rm{\AA\ pix^{-1}}$ in the band (\textit{Y} or \textit{J}) containing [\ion{O}{2}], 0.47 $\rm{\AA\ pix^{-1}}$ in the band (\textit{J} or \textit{H}) containing H$\gamma$ through [\ion{O}{3}] $\lambda$4959, and 0.63 $\rm{\AA\ pix^{-1}}$ in the band (\textit{H} or \textit{K}) containing H$\alpha$ and [\ion{N}{2}]. (Hereafter, these bands will be referred to as the \textit{YJ}-band, \textit{JH}-band, and \textit{HK}-band, respectively.) These rest-frame dispersions are computed by shifting the MOSFIRE \textit{J}-, \textit{H}-, and \textit{K}-band observed-frame dispersions$^{\rm{\ref{footnote:grating}}}$ to the median redshift of the stacking sample, $z_{\rm{med}}\approx2.465$. Once resampled, the spectra from the full galaxy sample are combined at each wavelength element by taking the median value of all luminosity densities at that point. When stacking the spectra of a multiply-imaged galaxy, the average at each pixel is taken instead, with luminosity densities weighted by their associated $1\sigma$ uncertainty values. Finally, the composite spectrum of the full stacking sample is multiplied by the median [\ion{O}{3}] $\lambda$5007 luminosity of the sample, whereas the composite for a multiply-imaged galaxy is multiplied by the lowest-luminosity [\ion{O}{3}] $\lambda$5007 measurement (a proxy for the least-magnified measurement). The final $1\sigma$ uncertainty spectrum for each multiple-image composite is the result of error propagation throughout the stacking process.  We discuss the construction of the full-sample composite uncertainty spectrum below. We note that, in this work, only ratios of emission lines are used from the composite of the full stacking sample since individual luminosity measurements rely on a normalization dependent on our stacking methodology.

\begin{figure*}[ht]
    \centering
    \includegraphics[trim={1cm 1.9cm 1.6cm 1.2cm}, width=\textwidth, clip]{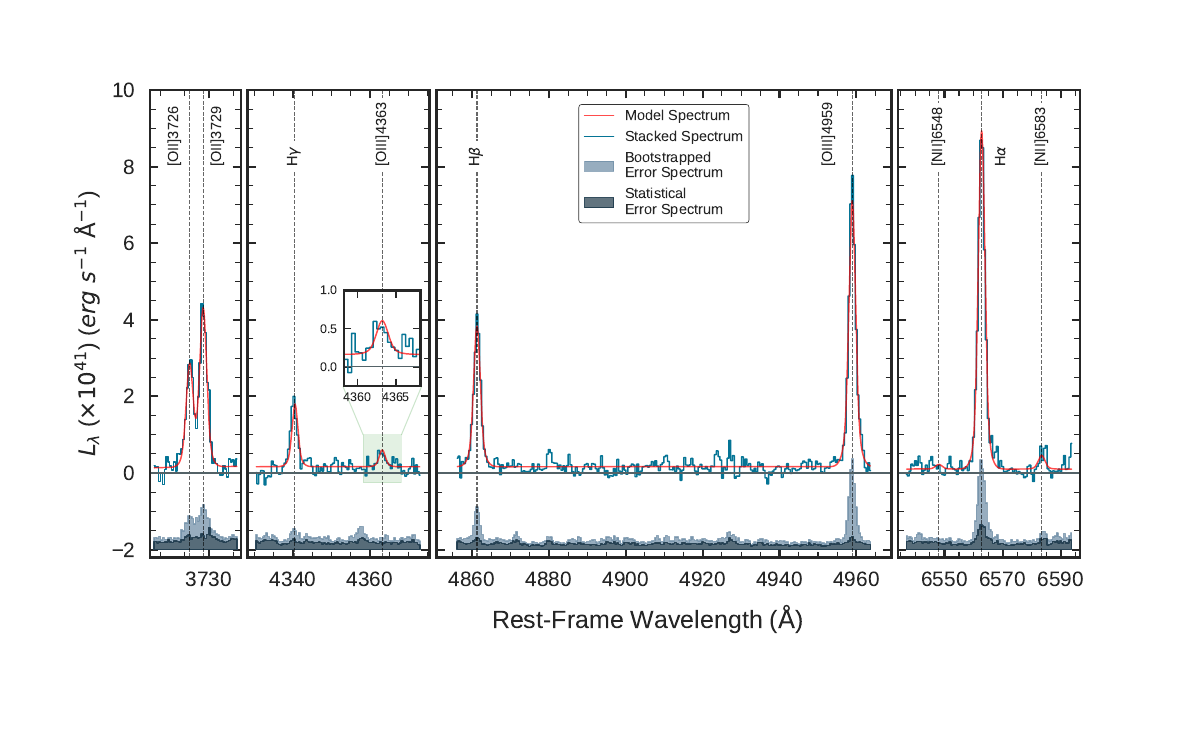}
    \caption{The rest-optical composite spectrum of the 16 typical, star-forming, $\langle z \rangle = 2.30$ dwarf galaxies in our stacking sample. The stacked spectrum is shown as the blue unfilled histogram, and its best-fit model is displayed in red. Here we show two offset uncertainty spectra (see Section \ref{subsec:stack}), one estimated via bootstrapping in order to account for sample variance (the light blue-gray-filled histogram), and the other a statistical-only error spectrum (the dark blue-gray-filled histogram). We note that the bootstrapped uncertainty spectrum was used when fitting the model shown here. In the inset box corresponding to the green-shaded spectral region, we show a zoom-in of the weak [\ion{O}{3}] $\lambda$4363 auroral line, measured at 2.5$\sigma$ (4.1$\sigma$) significance when considering the bootstrapped (statistical-only) uncertainty spectrum. We note that the luminosity density values ($L_\lambda$) of the composite have a constant but arbitrary normalization dependent on our stacking methodology (see Section \ref{subsec:stack}). While this makes individual emission-line luminosities unreliable, it does not affect line ratios. \newline \label{fig:comp_spectrum}}
\end{figure*}

The uncertainty spectrum of the stack of all of our sample galaxies is derived via a Monte Carlo approach with bootstrapping. We first create a bootstrapped sample of number count $N_{\rm{boot}}=16$, the number of galaxies in our full stack, by randomly drawing galaxies for the sample with replacement. For each galaxy in this bootstrapped sample, its science spectrum (already shifted and converted to the rest-frame and luminosity densities, respectively, during stacking above) is perturbed according to its $1\sigma$ luminosity density error spectrum. This perturbed spectrum is then normalized by its corresponding [\ion{O}{3}] $\lambda$5007 emission-line luminosity, which has also been perturbed based on its own uncertainty value. The normalized, perturbed spectra are then resampled and stacked according to the procedure detailed above. This process is repeated 500 times in order to create an array of composite luminosity densities at each wavelength element. The composite uncertainty spectrum is comprised of the standard deviations of the values in each of these arrays. By constructing our error spectrum via bootstrapping, our uncertainties represent both our measurement errors and sample variance. The composite spectrum of our full stacking sample, as well as its bootstrapped uncertainty spectrum (the light-gray-shaded region), are shown in Figure \ref{fig:comp_spectrum}. In Figure \ref{fig:comp_spectrum}, we also show the statistical-only uncertainty spectrum (the dark-gray-shaded region), which was created as described above, but without bootstrap-resampling each iteration. We note that, unless specified otherwise, the stated uncertainties on measurements derived from the composite spectrum reflect the use of the bootstrapped uncertainty spectrum and its consideration of sample variance.

\subsubsection{Fitting the Composite Emission-Line Spectrum} \label{subsubsec:fit_comp_spectra}
Prior to their inclusion in the full-sample stack, the composite spectra of the multiply-imaged galaxies are fit in the same manner as the spectra of individual galaxy images (see Section \ref{subsec:fit_ind_spectra}). For the full-sample composite, we adopt a slightly different fitting methodology. While spectral-fitting is still facilitated with \texttt{emcee}, and the continuum is still fit with a line, the emission lines are fit with a combination of two Gaussian profiles due to the non-Gaussian shape of the high-S/N lines$\,-\,$the deviation from a Gaussian profile likely resulting from the stacking itself \citep{Steidel2016}. Both of these Gaussian components are centered on the rest-frame wavelengths of the emission lines being fit. In order to maintain consistent resultant line profiles for all emission lines in a given spectral band (\textit{YJ}, \textit{JH}, or \textit{HK}), the fitting of these composite profiles, and spectral bands, is done in two rounds. We note that in fitting two Gaussian components to each emission line, we make no attempt to constrain any physical processes, such as outflows, that are often studied via decomposed emission lines. Any widths we fit are reported here but are not to be considered physical. Instead, our goal is simply to obtain more accurate fits to the emission lines of the composite spectrum.

In the first round of fitting, the two Gaussian components are fit to the brighter, higher-S/N lines in the composite: H$\beta$ and [\ion{O}{3}] $\lambda$4959 in the \textit{JH}-band and, separately, H$\alpha$ in the \textit{HK}-band. For each spectral band, one of the Gaussian components (hereafter referred to as the ``set-width" or ``SW" component) has its $1\sigma$-width set at $100\ \rm{km\ s^{-1}}$ while the other Gaussian component's (hereafter referred to as the ``free-width" or ``FW" component) $1\sigma$-width is left as a free parameter. In addition to the singular FW-component $1\sigma$-width, the free parameters for each fit include the slope and intercept of the continuum, the amplitude(s) of the FW profile(s), and a FW$\,-\,$to$\,-\,$SW amplitude ratio for the spectral band.

In the second round of fitting, each spectral band is fit in its entirety. The SW component maintains its $1\sigma$-width of $100\ \rm{km\ s^{-1}}$ in all bands. In the \textit{JH}- and \textit{HK}-bands, the $\sigma_{\rm{FW}}$ values and FW$\,-\,$to$\,-\,$SW amplitude ratios fit in the first round are held fixed and applied to all lines in the corresponding bands: H$\gamma$, [\ion{O}{3}] $\lambda$4363, H$\beta$, [\ion{O}{3}] $\lambda$4959 in the \textit{JH}-band and H$\alpha$ and [\ion{N}{2}] $\lambda\lambda$6548, 6583 in the \textit{HK}-band. Free parameters in these bands during this second round of fitting are the FW-component amplitudes of each line and the linear continuum parameters. (The [\ion{N}{2}] $\lambda$6583 FW amplitude is constrained here in the same manner as this line's amplitude in the individual spectra; see Section \ref{subsec:fit_ind_spectra}.) In the \textit{YJ}-band containing [\ion{O}{2}] $\lambda\lambda$3726, 3729, the amplitude ratio from the \textit{JH}-band is adopted, but $\sigma_{\rm{FW}}$ is left as a free parameter. Like with the other bands, the FW amplitudes and linear continuum parameters are also fit. In all bands, we find the best-fit $1\sigma$-widths of the FW Gaussian component to be $\sigma_{\rm{FW}}\approx50\ \rm{km\ s^{-1}}$. In the \textit{JH}- and \textit{HK}-bands, we find the best-fit FW$\,-\,$to$\,-\,$SW amplitude ratios to be 4.6 and 6.4, respectively.

The resultant spectral model from these two rounds of fitting can be seen in red in Figure \ref{fig:comp_spectrum}. The total luminosity of each emission line, representing the addition of the SW and FW Gaussian component luminosities, is given in Table \ref{tab:luminosities} relative to the total H$\beta$ luminosity. The reported total luminosities of the H$\gamma$, H$\beta$, and H$\alpha$ emission lines have been corrected for stellar absorption, reflecting an increase in the measured luminosities by the sample-median values of $\sim5.0\%$, $\sim1.7\%$, and $\sim0.4\%$, respectively (see Section \ref{subsec:fit_ind_spectra}). Similar to the emission-line flux uncertainties estimated in Section \ref{subsec:fit_ind_spectra}, uncertainties on the measured emission-line luminosities here are taken to be the $1\sigma$-widths of the total luminosity posterior distributions resulting from the fitting process. Of particular interest for this study is the $2.5\sigma$ ($4.1\sigma$) detection of [\ion{O}{3}] $\lambda$4363 in our composite with the bootstrapped (statistical-only) error spectrum, which will be used in Sections \ref{subsec:Te_ne} and \ref{subsec:metallicity} to estimate the composite electron temperature and gas-phase metallicity directly.\footnote{In low-redshift, high-metallicity galaxies ($12+\log(\rm{O/H})\geqslant8.3$), [\ion{O}{3}] $\lambda$4363 can be significantly contaminated by the [\ion{Fe}{2}] $\lambda$4359 emission line, resulting in the overestimation of $T_e$([\ion{O}{3}]) and the underestimation of O/H \citep[e.g.,][]{Curti2017}. Fortunately, as we show in Section \ref{subsec:metallicity}, our composite metallicity is $12+\log(\rm{O/H}) = 7.88$, well below the metallicities at which the Fe contamination is relevant. This, combined with the super-solar O/Fe abundance ratios seen in high-$z$ star-forming galaxies \citep[e.g.,][]{Steidel2016}, removes any concern of contamination of our [\ion{O}{3}] $\lambda$4363 line.}

We note that we tested the validity of assuming the SW-component's $1\sigma$-width by comparing luminosities of H$\beta$, [\ion{O}{3}] $\lambda$4959, and H$\alpha$ fit with either the SW-component's width set to $\sigma_{\rm{SW}} = 100\ \rm{km\ s^{-1}}$ or the width left as a free parameter (the rest of the fitting methodology unchanged). When comparing the luminosities fit assuming our fiducial $\sigma_{\rm{SW}}$ to the weighted-average luminosities of three runs where $\sigma_{\rm{SW}}$ was left free, we found an average percentage difference of $\sim 0.4\%$ and $\sim 1.0\%$ when assuming the bootstrapped or statistical-only uncertainty spectrum, respectively. The weighted-average luminosities fell well within the uncertainties of the luminosities measured when assuming $\sigma_{\rm{SW}}$. We therefore find our choice of $\sigma_{\rm{SW}} = 100\ \rm{km\ s^{-1}}$ to be robust.\footnote{When $\sigma_{\rm{SW}}$ was left free, we recovered best-fit values of $\sigma_{\rm{SW}} \approx 120-125\ \rm{km\ s^{-1}}$ and $\sigma_{\rm{SW}} \approx 100-115\ \rm{km\ s^{-1}}$ when fitting H$\beta$ and [\ion{O}{3}] $\lambda$4959, or H$\alpha$, respectively. While these $\sigma_{\rm{SW}}$ values are larger than the assumed value of $\sigma_{\rm{SW}} = 100\ \rm{km\ s^{-1}}$, so too are the corresponding best-fit FW$\,-\,$to$\,-\,$SW amplitude ratios, which are $\sim1.5-2\times$ larger when $\sigma_{\rm{SW}}$ is a free parameter. Unfortunately, when $\sigma_{\rm{SW}}$ is left free, there are some concerns as to how well-constrained the amplitude ratios are, particularly when fitting H$\alpha$, as well as concerns of potential over-fitting of the SW-component wings for H$\beta$ and [\ion{O}{3}] $\lambda$4959. We do not have these concerns when assuming $\sigma_{\rm{SW}}$. Of note, regardless of whether $\sigma_{\rm{SW}}$ was set or left free, $\sigma_{\rm{FW}}$ was consistently found to be $\sigma_{\rm{FW}}\approx 50\ \rm{km\ s^{-1}}$.}

\begin{deluxetable}{lcrrc}[t!]
\vspace{0.1cm}
\tablecaption{Emission-Line Luminosities of the Composite Spectrum \label{tab:luminosities}}
\tablecolumns{5}
\tablenum{2}
\tablewidth{\textwidth}
\renewcommand{\arraystretch}{1.2}
\setlength{\tabcolsep}{6.8pt}
\tablehead{
\colhead{Line} &
\colhead{$\lambda_{\rm rest}\tablenotemark{a}$} &
\colhead{$L_{\rm{meas}}\tablenotemark{b}$} &
\colhead{$L_{\rm{corr}}\tablenotemark{b,c}$} &
\colhead{$A_\lambda\tablenotemark{d}$}
}
\startdata
$[$\ion{O}{2}$]$ & 3726.032 & $0.61 \pm 0.12$ & $0.70 \pm 0.13$ & 0.67 \\
$[$\ion{O}{2}$]$ & 3728.815 & $0.95 \pm 0.16$ & $1.09 \pm 0.19$ & 0.67 \\
H$\gamma$        & 4340.459 & $0.40 \pm 0.08$ & $0.44 \pm 0.09$ & 0.59 \\
$[$\ion{O}{3}$]$ & 4363.209 & $0.11 \pm 0.05$ & $0.11 \pm 0.05$ & 0.58 \\
H$\beta$         & 4861.321 & $1.00 \pm 0.16$ & $1.00 \pm 0.16$ & 0.51 \\
$[$\ion{O}{3}$]$ & 4958.910 & $1.95 \pm 0.30$ & $1.89 \pm 0.29$ & 0.49 \\
H$\alpha$        & 6562.794 & $3.14 \pm 0.48$ & $2.70 \pm 0.42$ & 0.36 \\
$[$\ion{N}{2}$]$ & 6583.448 & $0.13 \pm 0.07$ & $0.11 \pm 0.06$ & 0.35 \\
\enddata
\tablecomments{The luminosity of [\ion{O}{3}] $\lambda$5007 can be calculated via the intrinsic ratio [\ion{O}{3}] $\lambda$5007/[\ion{O}{3}] $\lambda$4959 = 2.98.}
\tablenotetext{a}{Rest-frame wavelengths in air ($\rm{\AA}$).}
\tablenotetext{b}{Luminosities relative to $L_{\rm{H}\beta}$. We only use ratios of the luminosities from the composite spectrum since the spectrum's normalization is dependent on our stacking methodology.}
\tablenotetext{c}{All luminosities are dust-corrected using the corresponding extinction magnitude ($A_\lambda$) given in the last column (see Section \ref{subsec:dust_sfr}). Balmer lines are corrected for stellar absorption (see Section \ref{subsec:fit_ind_spectra}). The listed uncertainties do not include systematic errors associated with the dust correction, though these errors are propagated throughout all of our calculations.}
\tablenotetext{d}{Dust extinction magnitudes at $\lambda_{\rm{rest}}.$}
\vspace{-0.8cm}
\end{deluxetable}

\subsection{Dust Extinction and SFRs of the Sample} \label{subsec:dust_sfr}
In order to estimate a galaxy's intrinsic emission-line luminosities, from which its galactic properties and interstellar medium (ISM) conditions are derived, a wavelength-dependent correction to the observed luminosities must be made to account for extinction from nebular dust. This correction is typically quantified via observed hydrogen Balmer recombination-line ratios. Ideally, dust extinction would have been compensated for on a galaxy-by-galaxy basis prior to stacking our sample. Unfortunately, many of our individual galaxy spectra have Balmer lines that are too faint or too impacted by sky lines for this approach to be used. Instead, the nebular dust extinction ``typical" of star-forming, dwarf galaxies like those in our sample is estimated via our full-sample composite spectrum.

To calculate this typical nebular dust extinction, we utilized our stellar-absorption-corrected, composite hydrogen Balmer emission lines and assumed Case B intrinsic Balmer ratios of H$\alpha$/H$\beta$ = 2.79 and H$\alpha$/H$\gamma$ = 5.90 for an electron temperature and electron density of $T_e$([\ion{O}{3}]) = 15,000 K and $n_e$ = 100 $\rm{cm^{-3}}$, respectively\footnote{We note that while an electron temperature is assumed when selecting Balmer ratio values, the dependence of those ratios on $T_e$ is weak for typical temperatures in \ion{H}{2} regions. To confirm that our assumption of $T_e$ = 15,000 K is valid, we dust-corrected the composite spectrum then calculated $T_e$ (see Section \ref{subsec:Te_ne}) assuming intrinsic Balmer ratios corresponding to $T_e$ = 10,000 K, 12,500 K, 15,000 K, and 20,000 K \citep{Dopita&Sutherland2003}. With each variation, we consistently calculated from our dust-corrected composite spectrum a $T_e\sim\ $15,000 K.} \citep{Dopita&Sutherland2003}. Further assuming the extinction curve of \citet{Cardelli1989} with $R_V$ = 3.1, we find a ``typical" color excess of $E(\bv)_{\rm{gas}}=A_V/R_V=0.14^{+0.11}_{-0.09}$. This result allows us to correct our composite emission-line luminosities for extinction due to dust and to calculate the typical intrinsic emission-line ratios of star-forming, dwarf galaxies at high redshift. The dust-corrected (and stellar-absorption-corrected in the case of the Balmer lines) emission-line luminosities of the composite spectrum, relative to H$\beta$, are listed in Table \ref{tab:luminosities}. We note that when later calculating typical intrinsic strong-line ratios and physical properties of our stack, we first randomly and independently sample ($N = 100,000$) $A_V$ and the observed emission-line luminosities based on their respective probability distributions. This subsequently gives us samples of dust-corrected (on an element-by-element basis) emission-line luminosities with which we make our calculations. The best-fit values and uncertainties of the ratios and properties are taken to be the most probable values and $68\%$ confidence intervals, respectively, of the corresponding posteriors. 

When considering SFRs, similar to how our stellar masses are being reported, we calculate the SFR for each individual galaxy and report the composite SFR as the median value of the sample. These SFRs are calculated with slit-loss-corrected, dust-corrected, H$\alpha$ luminosities ($L_{\rm{H\alpha}}$), de-magnified according to the lensing models in Section \ref{subsubsec:lensmodels}. The dust-extinction correction of each $L_{\rm{H}\alpha}$ value is conducted with the ``typical" extinction estimate for the sample found via the composite spectrum. The $L_{\rm{H}\alpha}$ values are not corrected for stellar absorption, which on average would result in an increase of $< 1\%$. The $L_{\rm{H\alpha}}$ values are converted to SFRs using Equation \ref{equ:SFR} below:

\begin{equation} \label{equ:SFR}
    \mathrm{SFR}\ (\mathrm{M_\odot\ yr^{-1}}) = 4.645\times10^{-42}\ L_{\mathrm{H}\alpha}\ (\mathrm{ergs\ s^{-1}})
\end{equation}

\noindent This equation is of the same form as the relation in \citet{Kennicutt1998} for calculating SFRs from recombination lines. However, the conversion factor here has been recalculated assuming a metallicity of 0.2 $\rm{Z_\odot}$ and a \citet{Chabrier2003} IMF. Our estimates for the SFRs of the individual galaxies are given in Table \ref{tab:stacking_sample}. The SFR associated with the composite spectrum is taken to be the sample-median value of $\rm{SFR_{med}} = 2.25^{+2.15}_{-1.26}\ \rm{M_\odot\, yr^{-1}}$, with the stated error bars representing the interquartile range (IQR) of the SFRs. 

\subsection{Electron Temperature and Electron Density} \label{subsec:Te_ne}
The ``direct" calculation of metallicity relies on collisionally-excited oxygen emission lines and the nebular properties of electron temperature ($T_e$) and electron density ($n_e$), which are responsible for the strength of the collisionally-excited lines. Electron temperature is calculated in two ionization zones of the star-forming, \ion{H}{2} regions. In the $\rm{O^{++}}$ zone, $T_e$([\ion{O}{3}]) is calculated using the electron-temperature-sensitive emission-line ratio [\ion{O}{3}] $\lambda\lambda$4959, 5007/[\ion{O}{3}] $\lambda$4363 and the \texttt{getTemDen} method (with the default [\ion{O}{2}] and [\ion{O}{3}] atomic data) of the \texttt{PyNeb}\footnote{\url{https://pypi.org/project/PyNeb/} (version: 1.1.16)}$^{,}$\footnote{\url{http://morisset.github.io/PyNeb_devel/}}$^{,}$\footnote{\url{https://github.com/Morisset/PyNeb_devel/tree/master/docs}} emission-line analysis software \citep{Luridiana2015}. We note that while $T_e$([\ion{O}{3}]) does have a dependence on electron density, $n_e$, below $n_e\approx10^3\ \rm{cm^{-3}}$, $T_e$([\ion{O}{3}]) is insensitive to $n_e$ \citep{Osterbrock_Ferland2006,Izotov2006} and can be calculated assuming the typical $z\sim2$ \ion{H}{2} region electron density of a few hundred per cubic centimeter \citep{Sanders2016ne_u}. For our calculation, we assume $n_e=150\ \rm{cm^{-3}}$ and obtain an electron temperature in the $\rm{O^{++}}$ region of $T_e$([\ion{O}{3}]) = 15,500 $\pm$ 3,100 K. We note in regard to the assumed $n_e$ that \citet{Gburek2019} studied a galaxy in our present stacking sample, A1689-217, that had a similar electron temperature of $T_e$([\ion{O}{3}]) = 14,300 K. When calculating this temperature, they found that assuming any $n_e<10^3\ \rm{cm^{-3}}$ changed their result by $<0.5\%$, suggesting our current assumption is robust.

Ideally, the electron temperature in the $\rm{O^+}$ ionization region is calculated using measurements of the [\ion{O}{2}] $\lambda\lambda$7320, 7330 auroral emission-line doublet. Unfortunately, for the galaxies in our stacking sample, we do not have spectroscopic coverage of these lines. Instead, we calculate $T_e$([\ion{O}{2}]) via the $T_e$([\ion{O}{3}])$-T_e$([\ion{O}{2}]) relation of \citet{Campbell1986}, reprinted here in Equation \ref{equ:teo3_teo2}:

\begin{equation} \label{equ:teo3_teo2}
    T_e\text{([\ion{O}{2}])} = 0.7\, T_e\text{([\ion{O}{3}])} + 3000\ \text{K}
\end{equation}

\noindent Use of this equation gives us an electron temperature in the $\rm{O^+}$ region of $T_e$([\ion{O}{2}]) = 13,900 $\pm$ 2,100 K.

The electron density, $n_e$, can be derived with the doublet ratio [\ion{O}{2}] $\lambda$3729/[\ion{O}{2}] $\lambda$3726, the $T_e$([\ion{O}{2}]) electron temperature, and the \texttt{getTemDen} method of the \texttt{PyNeb} software. For our composite spectrum, we calculate [\ion{O}{2}] $\lambda$3729/[\ion{O}{2}] $\lambda$3726 = 1.56 $\pm$ 0.32 (1.51 $\pm$ 0.12 when using the statistical-only uncertainty spectrum and associated fits). This corresponds to an electron density of $n_e = 1^{+215}_{-0}\ \rm{cm^{-3}}$ ($n_e = 1^{+74}_{-0}\ \rm{cm^{-3}}$), where the ``best-fit" $n_e$ value is set to the low-density limit of $n_e = 1\ \rm{cm^{-3}}$ as a result of the best-fit [\ion{O}{2}] ratio exceeding the maximum theoretical bound of [\ion{O}{2}] $\lambda$3729/[\ion{O}{2}] $\lambda$3726 $\lesssim1.5$ \citep{Osterbrock_Ferland2006,Sanders2016ne_u}. 

We note that we have significantly detected the component emission lines of the [\ion{O}{2}] $\lambda\lambda$3726, 3729 doublet and resolved their individual peaks (S/N(3726, 3729) = (6.4, 7.8); see Figure \ref{fig:comp_spectrum}). The electron density associated with our dwarf galaxy sample and with the best-fit ratio of these lines is significantly lower than the densities found in more massive galaxies at $1.5 \lesssim z \lesssim 2.5$, which typically lie in the range of $n_e \approx 100-300\ \rm{cm^{-3}}$ \citep{Steidel2014, Sanders2016ne_u, Kashino2017, Kaasinen2017, Davies2021}.

\subsection{Direct Oxygen Abundance} \label{subsec:metallicity}
We directly calculate the oxygen abundance, or gas-phase metallicity, of our composite spectrum using the ionic abundance equations of \citet{Izotov2006}. These equations utilize the values of $T_e$([\ion{O}{2}]), $T_e$([\ion{O}{3}]), and $n_e$ given in the preceding section as well as the dust-corrected emission-line ratios of [\ion{O}{2}] $\lambda\lambda$3726, 3729/H$\beta$ and [\ion{O}{3}] $\lambda\lambda$4959, 5007/H$\beta$. We assume that the total oxygen abundance is the summation of the ionic abundances in the \ion{H}{2} region $\rm{O^+}$ and $\rm{O^{++}}$ ionization zones as seen in Equation \ref{equ:oxygen_abundance}. Any higher ionization states of oxygen are deemed to have a negligible contribution to the metallicity. 

\begin{equation} \label{equ:oxygen_abundance}
    \mathrm{\frac{O}{H}\ \approx\ \frac{O^+}{H^+} + \frac{O^{++}}{H^+}}
\end{equation}

From our composite spectrum, we report a typical gas-phase metallicity for high-redshift ($z \sim 2.3$), star-forming, dwarf galaxies of $12+\log(\rm{O/H}) = 7.88^{+0.25}_{-0.22}$ \citep[$0.15^{+0.12}_{-0.06}\ \rm{Z_\odot}$;][]{Asplund2021}. This metallicity estimate, as well as the calculations from our composite spectrum of the other physical properties detailed in Section \ref{sec:measurements}, are summarized in Table \ref{tab:comp_properties}. As a reminder, unless specified otherwise, the measurements presented in the text and in Table \ref{tab:comp_properties} have stated uncertainty values reflecting our use of the bootstrapped uncertainty spectrum of our composite and its consideration of sample variance.

\begin{deluxetable}{lc}[ht!]
\tablecaption{Properties of the Dwarf Galaxy Composite \label{tab:comp_properties}}
\tablecolumns{2}
\tablenum{3}
\setlength{\tabcolsep}{25pt}
\renewcommand{\arraystretch}{1.2}
\tablewidth{\textwidth}
\tablehead{
\colhead{Property} &
\colhead{Value}
}
\startdata
$z_{\rm{mean}}\tablenotemark{a}$ & 2.30 \\
$\log(M_\ast/\rm{M_\odot})^{\rm{med}}_{\rm{fiducial}}\tablenotemark{a,b}$ & $8.29^{+0.51}_{-0.43}$ \\
$\log(M_\ast/\rm{M_\odot})_{\rm{med}}\tablenotemark{a,c}$ & $8.92^{+0.31}_{-0.22}$ \\
$\rm{SFR}^{\rm{med}}_{\rm{H}\alpha}\ (\rm{M_\odot\ yr^{-1}})\tablenotemark{a}$ & $2.25^{+2.15}_{-1.26}$ \\
$E(\bv)_{\rm{gas}}$ & $0.14^{+0.11}_{-0.09}$ \\
$n_e^{\rm{boot}}\ (\rm{cm^{-3}})\tablenotemark{d}$ & $1^{+215}_{-0}$ \\
$n_e^{\rm{stat}}\ (\rm{cm^{-3}})\tablenotemark{d}$ & $1^{+74}_{-0}$ \\
$T_e$([\ion{O}{2}]) (K) & 13,900 $\pm$ 2,100 \\
$T_e$([\ion{O}{3}]) (K) & 15,500 $\pm$ 3,100 \\
$12+\log(\rm{O^+/H^+})$ & $7.30^{+0.26}_{-0.20}$ \\
$12+\log(\rm{O^{++}/H^+})$ & $7.75^{+0.24}_{-0.23}$ \\
$12+\log(\rm{O/H})_{\rm{direct}}$ & $7.88^{+0.25}_{-0.22}$ \\
$Z\ (\rm{Z_\odot})$ & $0.15^{+0.12}_{-0.06}$
\enddata
\tablecomments{All uncertainties here (except for $n_e^{\rm{stat}}$ and the interquartile ranges (IQR) reported with the median values) derive from the composite bootstrapped error spectrum. See Figure \ref{fig:comp_spectrum}.}
\tablenotetext{a}{Mean and median values of the individual galaxies in the stacking sample. Median values are reported with the interquartile range (IQR) of the corresponding property. See Table \ref{tab:stacking_sample}. All other values derive from the composite spectrum.}
\tablenotetext{b}{Our fiducial median stellar mass assuming constant SFHs.}
\tablenotetext{c}{The median stellar mass assuming non-parametric SFHs. See Section \ref{subsubsec:sim_mzrs}.}
\tablenotetext{d}{$n_e^{\rm{boot}}$ and $n_e^{\rm{stat}}$ assume the bootstrapped and statistical-only error spectrum (and associated fits), respectively. Both ``best-fit" values are set as the low-density limit of $n_e = 1\ \rm{cm^{-3}}$. See Section \ref{subsec:Te_ne}.}
\vspace{-0.8cm}
\end{deluxetable}

\section{Results and Discussion} \label{sec:results_discussion}
This section of the manuscript will take the measurements derived in the previous section from our dwarf galaxy sample and composite spectrum and analyze them in the context of strong-line abundance diagnostics and global galaxy scaling relations. Prior to this, however, it is crucial to look at our sample and stack compared to the broader star-forming galaxy population at $1.7 \lesssim z \lesssim 2.6$ in order to assess how representative our sample is, on average, of typical dwarf galaxies at this epoch. In this forthcoming comparison, and presentation and discussion of our results, we emphasize that our findings are based on our composite and sample-median values and therefore are applicable to $z \sim 2.3$ star-forming dwarf galaxy samples \textit{on average}. Our results may not accurately determine or reflect the physical properties in individual high-$z$ dwarf galaxies due to the intrinsic variation of properties from galaxy-to-galaxy. 

\subsection{How Representative is our Sample?}\label{subsec:representative}
Here we will consider two main diagnostics, the [\ion{N}{2}] Baldwin-Phillips-Terlevich diagram \citep[N2-BPT;][]{BPT1981} and the $M_\ast-\rm{SFR}$ relation, or ``star-forming main sequence."

\begin{figure}[ht!]
    \includegraphics[trim={0.3cm 0.5cm 0.1cm 0cm}, width=\columnwidth, clip]{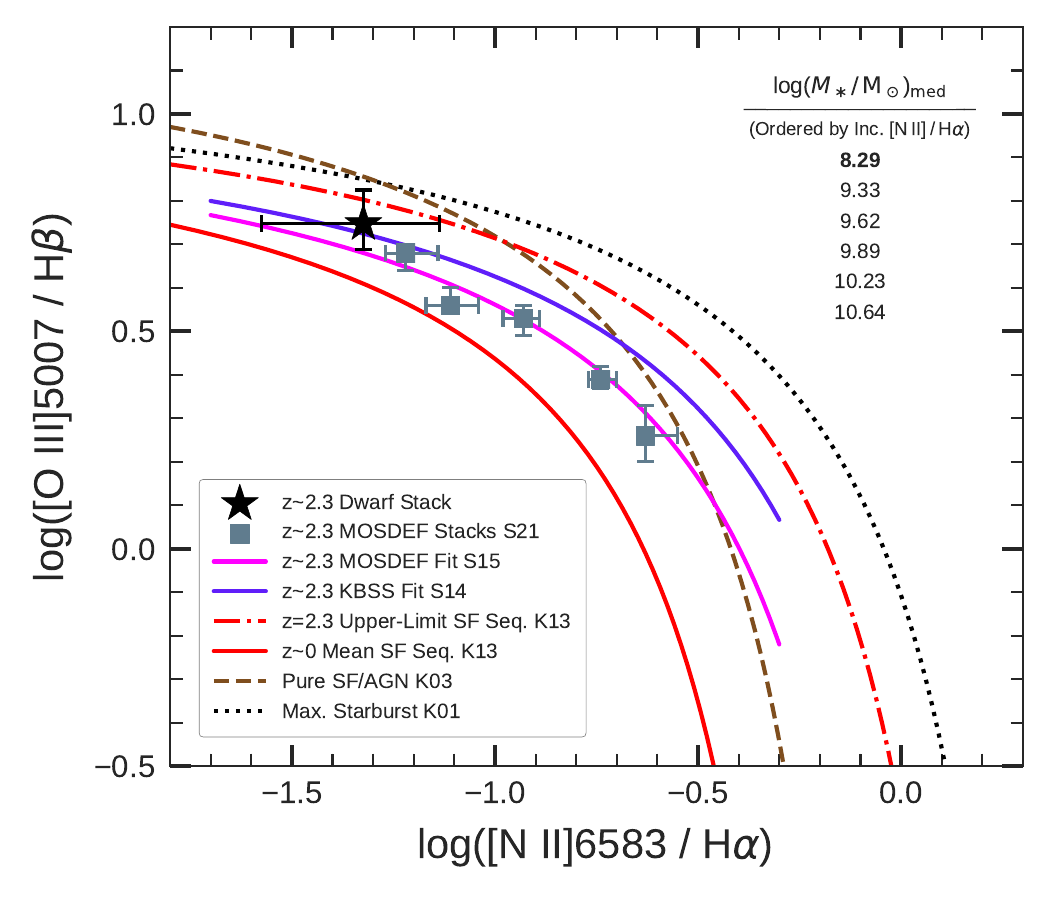}
    \caption{The [\ion{O}{3}] $\lambda$5007/H$\beta$ vs. [\ion{N}{2}] $\lambda$6583/H$\alpha$ BPT diagnostic diagram. Our $\langle z \rangle = 2.30$ dwarf galaxy composite is shown as the black star. The $z \sim 2.3$ $M_\ast$-binned stacks of star-forming (SF) MOSDEF galaxies from \citet[][S21]{Sanders2021} are shown as gray squares. The median $M_\ast$ of each of these stacks, and of our composite (in bold), is listed in the upper right-hand corner of the plot by order of increasing [\ion{N}{2}]/H$\alpha$, highlighting that our stack is an extension to lower $M_\ast$ (and O/H via the MZR) of the MOSDEF survey. The $z \sim 2.3$ SF sequences of the MOSDEF \citep[][S15]{Shapley2015} and KBSS-MOSFIRE \citep[][S14]{Steidel2014} surveys are shown by the magenta and purple lines, respectively. Like these sequences and the MOSDEF stacks, our composite also lies offset from the $z \sim 0$ mean SF sequence given by the red line and parameterized by \citet[][K13]{Kewley2013}. We plot the $z=2.30$ SF sequence upper-limit from K13 as the red dot-dashed line. The demarcation between SF galaxies and AGN of \citet[][K03]{Kauffmann2003} is given by the dashed brown line, and the ``maximum starburst" curve of \citet[][K01]{Kewley2001} is given by the dotted black line. \newline \label{fig:BPT_stack}}
\end{figure}

\subsubsection{N2-BPT Diagnostic Diagram}\label{subsubsec:n2_bpt}
In Figure \ref{fig:BPT_stack}, we show the location of our $\langle z \rangle = 2.30$ stack of star-forming, dwarf galaxies on the [\ion{O}{3}] $\lambda$5007/H$\beta$ vs. [\ion{N}{2}] $\lambda$6583/H$\alpha$ BPT diagnostic diagram. This diagram is a useful tool for distinguishing between star-forming galaxies (SFGs) and AGN through optical strong-line ratios \citep[][K01, K03, K13, respectively]{Kewley2001, Kauffmann2003, Kewley2013}, all without needing to apply a dust correction to the line fluxes. The SFG locus of the BPT is also a probe of changing physical conditions in star-forming regions with redshift \citep[e.g.,][Figure 2]{Kewley2013}. This has been an active area of research in numerous high-redshift statistical studies \citep[e.g.,][S14, S15, respectively]{Steidel2014, Shapley2015} which have shown that high-$z$ SFGs cluster around a locus offset toward higher [\ion{O}{3}] $\lambda$5007/H$\beta$ and/or [\ion{N}{2}] $\lambda$6583/H$\alpha$ when compared to the star-forming locus of $z \sim 0$ SDSS galaxies. We show this in Figure \ref{fig:BPT_stack}, where the SFG locus of $z \sim 0$ SDSS galaxies is given by the solid red line \citep{Kewley2013}, and the offset SFG loci of the $z \sim 2.3$ KBSS-MOSFIRE and MOSDEF surveys are displayed by the purple \citep{Steidel2014} and magenta \citep{Shapley2015} lines, respectively. Recent work by \citet{Runco2022} has shown that these high-$z$ SFG loci actually converge with consistent emission-line-fitting applied to each sample. These authors' results also reaffirm the existence of an offset in the BPT between local and high-$z$ SFGs.

We show, via the black star in Figure \ref{fig:BPT_stack}, that our stack of $z \sim 2.3$ star-forming galaxies is also offset from the SDSS SFG locus, lying in parameter space consistent with KBSS-MOSFIRE and MOSDEF.\footnote{We note that the location of our stack on the BPT can also be occupied by low-metallicity fast shocks which can present in our spectroscopy as a faint, broad emission component \citep[e.g.,][Figure 11]{Allen2008, Kewley2019}. Measurement and analysis of this possible contaminant are beyond the scope of this paper, but we refer the reader to \citet{Freeman2019} for a study of broad nebular emission in higher-mass MOSDEF galaxies.} While we cannot place each individual galaxy in our sample on this plot due to skyline contamination, particularly of H$\beta$, we note that our stack lies below the ``maximum starburst" demarcation (dotted black line) of \citet{Kewley2001}, the empirical demarcation (dashed brown line) between SFGs and AGN of \citet{Kauffmann2003}, and the theoretical, $z=2.30$, upper-limit SFG locus (red dotted-dashed line) of \citet{Kewley2013}. We also note that the large uncertainty of our stack in log([\ion{N}{2}]/H$\alpha$) is the result of [\ion{N}{2}] $\lambda$6583 only being detected in the composite spectrum with 2$\sigma$ significance.

While the [\ion{N}{2}] $\lambda$6583 measurement in our composite is fairly uncertain, the location of our stack along the $x$-axis of the N2-BPT is interesting when compared to the $M_\ast$-binned stacks of SFGs (the blue-gray squares) from the MOSDEF survey and \citet[][S21]{Sanders2021}. This is because of the monotonic relationship that exists between log([\ion{N}{2}] $\lambda$6583/H$\alpha$) and metallicity; as this strong-line ratio increases, metallicity increases \citep{Pettini&Pagel2004, Maiolino2008, Curti2017, Bian2018, Sanders2021}. By the mass-metallicity relation \citep[e.g.,][]{Tremonti2004, Sanders2021}, as this ratio increases, the stellar mass of galaxies should then also increase on average. We see this with the MOSDEF stacks, where in the upper right-hand section of the plot, we list the median stellar masses of the stacks in order of increasing [\ion{N}{2}]/H$\alpha$. Here in this list we have also included, in bold, the median stellar mass of our dwarf galaxy stack, which, based on its positioning in the BPT, predictably has the lowest listed median stellar mass. In this, we show that our sample is a complementary extension in stellar mass to the MOSDEF (and KBSS-MOSFIRE) survey, extending its mass range into the dwarf galaxy regime ($M_\ast < 10^9\ \rm{M_\odot}$). 

\subsubsection{The Star-Forming Main Sequence}\label{subsubsec:SFMS}
One of the primary goals of this study is to analyze our dwarf galaxy composite relative to the $z \sim 2.3$ mass-metallicity relation (MZR) of star-forming galaxies (see Section \ref{subsec:z2.3_mzr}), which is a scaling relation between galaxy stellar mass ($M_\ast$) and gas-phase oxygen abundance, or metallicity (O/H). However, in order to properly contextualize our findings in relation to the broader $z \sim 2.3$ dwarf galaxy population, we must consider the SFRs associated with the stack and the sample that comprises it. This is due to the existence of the fundamental metallicity relation (FMR) between $M_\ast$, SFR, and O/H which has been demonstrated locally at $z \sim 0$ \citep[e.g.,][]{Mannucci2010, Mannucci2011, Lara-Lopez2010, Andrews&Martini2013, Curti2020} and at high redshift out to $z \sim 3.3$ \citep[e.g.,][]{Henry2013_june, Henry2013_oct, Cresci2019, Sanders2018, Sanders2021}. The FMR demonstrates that, at a fixed $M_\ast$, a galaxy with an above-average (below-average) SFR will typically have a below-average (above-average) metallicity. Therefore, if our stacking sample (and thus composite spectrum) is biased in SFR, it will not have an average metallicity representative of typical dwarf galaxies at $z \sim 2.3$. This would be problematic when comparing the metallicity of our composite to the low-mass end of the MZR.

\begin{figure}[t]
    \includegraphics[trim={0.3cm 0.5cm 0.1cm 0cm}, width=\columnwidth, clip]{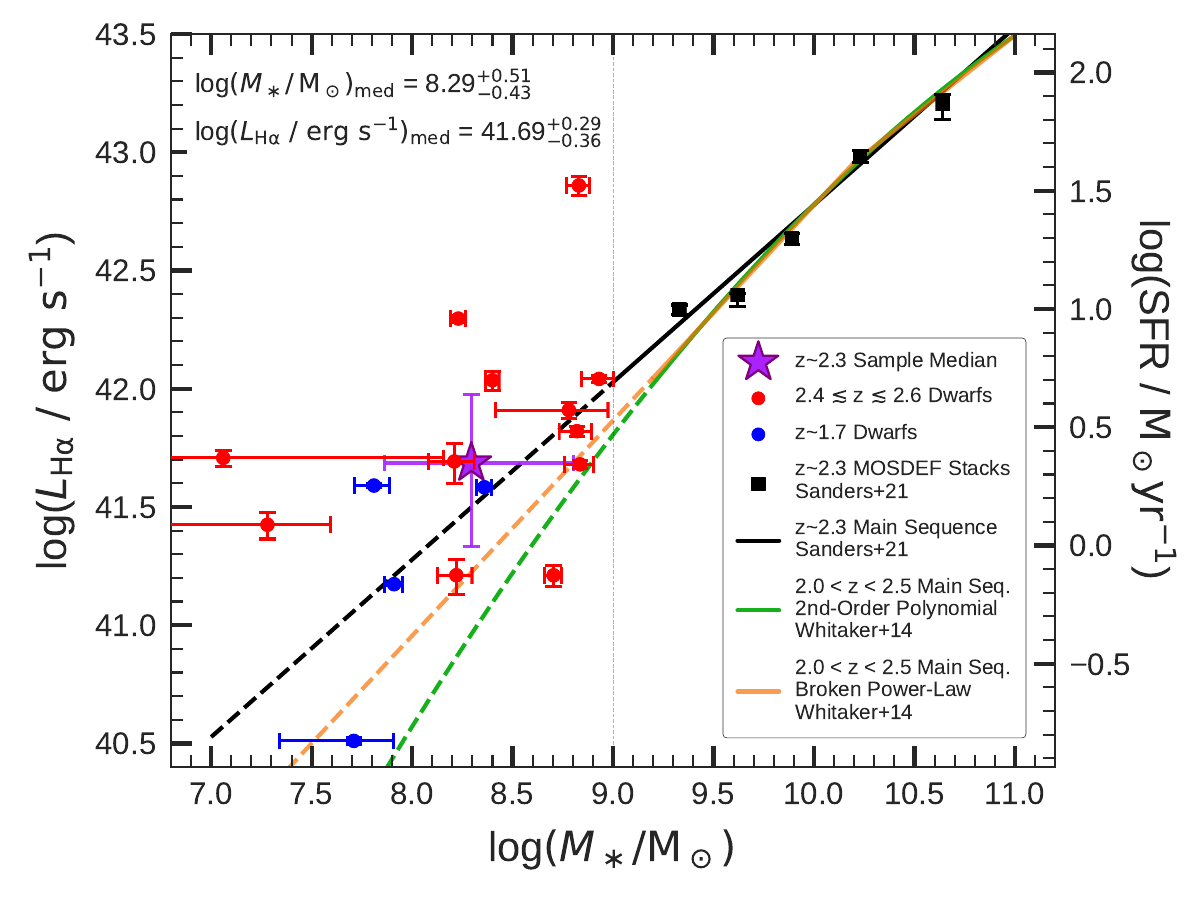}
    \caption{$M_\ast$ vs. dust-corrected $L_{\rm{H}\alpha}$ on the left-hand axis and SFR on the right-hand axis. Our stacking sample is shown by the colored circles, with blue circles representing our four $z \sim 1.7$ galaxies and red circles our twelve $2.4 \lesssim z \lesssim 2.6$ galaxies. The purple star lies at the median $M_\ast$ and $L_{\rm{H}\alpha}$ (or SFR) of our $z \sim 2.3$ sample. The error bars on the purple star represent the interquartile ranges of $M_\ast$ and $L_{\rm{H}\alpha}$ (or SFR) stated in the top-left corner. The black squares show the $z \sim 2.3$ $M_\ast$-binned stacks of MOSDEF galaxies from \citet{Sanders2021}. We compare our sample-median values against the $M_\ast-\rm{SFR}$ relations, or ``star-forming main sequences" (SFMS), of \citet[][black line]{Sanders2021} and \citet[][orange and green lines]{Whitaker2014} to determine how representative our sample is of typical, $z \sim 2.3$ dwarf galaxies. The \citet{Sanders2021} relation is a power-law derived from H$\alpha$ SFRs. The orange and green relations of \citet{Whitaker2014} are parameterized by a broken power-law and second-order polynomial, respectively; both relations derive from UV+FIR SFRs. Dashed portions of the SFMSs are extrapolations. \newline \label{fig:SFMS}}
\end{figure}

To investigate whether our stacking sample has a bias in SFR, we plot our sample and its median values against the $M_\ast-\rm{SFR}$ star-forming main sequence (SFMS) in Figure \ref{fig:SFMS}. Galaxies that lie on this mean relation, which is redshift-dependent, are considered to be representative of the typical galaxy at that corresponding stellar mass and redshift. In Figure \ref{fig:SFMS}, we compare to the $z \sim 2.3$ SFMS parameterizations of \citet{Sanders2021} and \citet{Whitaker2014}.

In comparing to the $z \sim 2.3$ SFMS of \citet{Sanders2021}, we do so with stellar masses and SFRs calculated in a manner highly consistent with the methodologies adopted in Sanders et al. The stellar masses of our stacking sample and the MOSDEF galaxies used to calibrate the SFMS both rely on emission-line-corrected photometry and the SED-fitting \citep[with \texttt{FAST};][]{Kriek2009} assumptions of constant star formation histories, the \citet{Calzetti2000} attenuation curve, and the \citet{Chabrier2003} IMF. The SFRs in both studies are calculated from dust-corrected (via the \citealt{Cardelli1989} extinction curve) H$\alpha$ luminosities ($L_{\rm{H}\alpha}$). We note that the conversion factor between $L_{\rm{H}\alpha}$ and SFR (see, for example, Equation \ref{equ:SFR}) is dependent on assumptions such as the stellar metallicity and can vary between different authors. Therefore, we plot dust-corrected $L_{\rm{H}\alpha}$ on the left-hand axis of Figure \ref{fig:SFMS} so that our stacking sample (red and blue circles) and the \citet{Sanders2021} MOSDEF stacks (black squares) can be directly compared without the additional SFR conversion. Ultimately, however, we find that the SFR conversion factor used by both studies is very similar, and we continue our analysis of how representative our stacking sample is via SFR, given on the right-hand axis of Figure \ref{fig:SFMS}.

The $z \sim 2.3$ SFMS of \citet{Sanders2021} is parameterized as a power-law over the stellar mass range 9.0 $<$ log($M_\ast$/$\rm{M_\odot}$) $<$ 11.0. In Figure \ref{fig:SFMS}, we plot their best-fit relation\footnote{The fitting of the SFMS was done with the four lowest-mass bins. Additionally, there is evidence that the lowest-mass ($M_\ast < 10^{9.5}\ \rm{M_\odot}$) bin may be biased high in SFR \citep{Shivaei2015, Sanders2021}.} over this range as a solid black line and extrapolate into the dwarf galaxy regime as seen by the dashed black line. In plotting our stacking sample in the $M_\ast-\rm{SFR}$ parameter space, we differentiate the $z \sim 1.7$ galaxies from the $2.4 \lesssim z \lesssim 2.6$ galaxies by blue and red points, respectively. We note that three of the four $z \sim 1.7$ galaxies lie along or above the extrapolation of the $z \sim 2.3$ SFMS. These objects are likely biased high in SFR relative to typical $z \sim 1.7$ galaxies that have a lower SFR at fixed $M_\ast$ due to the redshift evolution of the SFMS \citep{Speagle2014, Whitaker2014, Sanders2021}. The other group of $z > 2.38$ galaxies are found to scatter around the Sanders et al. extrapolation. Considered together, while our sample does have a large range in $M_\ast$ and SFR ($\sim 2$ orders of magnitude in each property), at the median mass of the complete $\langle z \rangle = 2.3$ stacking sample, log($M_\ast$/$\rm{M_\odot}$)$_{\rm{med}}$ = $8.29^{+0.51}_{-0.43}$, the median SFR of the sample, log(SFR/$\rm{M_\odot\,yr^{-1}}$)$_{\rm{med}} = 0.35^{+0.29}_{-0.36}$, lies only $\Delta \rm{log(SFR)} \approx 0.19$ dex above the extrapolation of the $z \sim 2.3$ SFMS of \citet{Sanders2021}. This median point is shown as the purple star, with its error bars representing the interquartile ranges of $M_\ast$ and SFR (or $L_{\rm{H}\alpha}$). 

With the relative offset in SFR of our stacking sample in hand, we estimate the bias in O/H, resulting from the FMR, of our composite spectrum. Considering the strength of the SFR-dependence of direct-method O/H at fixed $M_\ast$ from \citet{Sanders2020},

\begin{equation} \label{equ:o/h_sfr_relation}
    \Delta \rm{log(O/H)} \approx -0.29 \times \Delta \rm{log(SFR\,/\,M_\odot\, yr^{-1})}
\end{equation}

\noindent our sample stack is biased by $\Delta \rm{log(O/H)} \approx -0.06$ dex, a value half the statistical uncertainty of our composite direct-method metallicity estimate ($\sigma_{\rm{stat}} \approx 0.12$ dex). We therefore conclude that, when comparing to the SFMS of \citet{Sanders2021}, our stacking sample of dwarf galaxies does not have a major bias in SFR or O/H on average and, on average, is representative of typical dwarf galaxies at $z \sim 2.3$ with $M_\ast \gtrsim 10^8\ \rm{M_\odot}$.

In Figure \ref{fig:SFMS}, we also plot the $2.0 < z < 2.5$ SFMS parameterizations of \citet{Whitaker2014}, which were fit to $M_\ast$-binned stacks above a mass-completeness limit of $10^{9.2}\ \rm{M_\odot}$. Whereas \citet{Sanders2021} fit the SFMS with a power-law, \citet{Whitaker2014} fit the SFMS with both a second-order polynomial (green line) and a broken power-law (orange line) for which a separate slope was fit above and below a characteristic mass of log($M_\ast$/$\rm{M_\odot}$) = 10.2. We note that in our recreation of the second-order polynomial fit, we use the more precise polynomial coefficients given in the erratum \citep{Whitaker2020} to \citet[][Table 1]{Whitaker2014}, in order to more accurately portray the curve. Similar to \citet{Sanders2021} and the MOSDEF survey, the sample of \citet{Whitaker2014} is composed of star-forming galaxies from the CANDELS fields \citep{Grogin2011, Koekemoer2011}, though has a larger galaxy count and different sample selection. Like with the MOSDEF galaxies and our stacking sample, the stellar masses are determined with \texttt{FAST} assuming a \citet{Calzetti2000} attenuation curve and \citet{Chabrier2003} IMF; however, the star-formation histories are taken to be exponentially declining. Unlike in \citet{Sanders2021} and our stacking sample though, the SFRs are estimated from the combination of rest-frame ultraviolet (UV) light and light re-radiated by dust in the far-infrared (FIR). 

Figure \ref{fig:SFMS} shows that, at least qualitatively, the parameterizations from \citet{Whitaker2014} generally agree with the power-law fit ($\beta = 0.75$) and MOSDEF stacks of \citet{Sanders2021}, though begin to diverge as a result of steeper slopes near unity below log($M_\ast$/$\rm{M_\odot}$) $\lesssim 10$. This divergence, in part due to the choices of parametric form made by each author, is particularly pronounced in the dwarf galaxy mass regime where our sample lies. Unfortunately, there does not currently exist a high-redshift statistical sample in this regime that would allow us to confidently use a given parameterization. Therefore, we determine the most applicable SFMS based on consistency with our study in physical property estimation as well as the selection approach of our stacking sample. As detailed above, our methodologies for estimating $M_\ast$ and SFR are most analogous to those of \citet{Sanders2021}, mitigating systematic uncertainties between estimations of physical properties calculated with different techniques. Additionally, when selecting our stacking sample, we did not require a detection of [\ion{O}{3}] $\lambda$5007, H$\alpha$, or [\ion{O}{3}] $\lambda$4363 which typically bias a sample toward higher SFR at fixed $M_\ast$. Our sample selection instead suggests a more representative sample like we see when comparing our sample median to the SFMS of \citet{Sanders2021}. We therefore conclude that the SFMS of \citet{Sanders2021} is the most applicable comparison and that our stacking sample, on average, is representative of typical, star-forming, dwarf galaxies at $z \sim 2.3$.

\begin{figure*}[ht]
    \centering
    \includegraphics[trim={1.2cm 0.6cm 2.2cm 0.3cm}, width=\textwidth, clip]{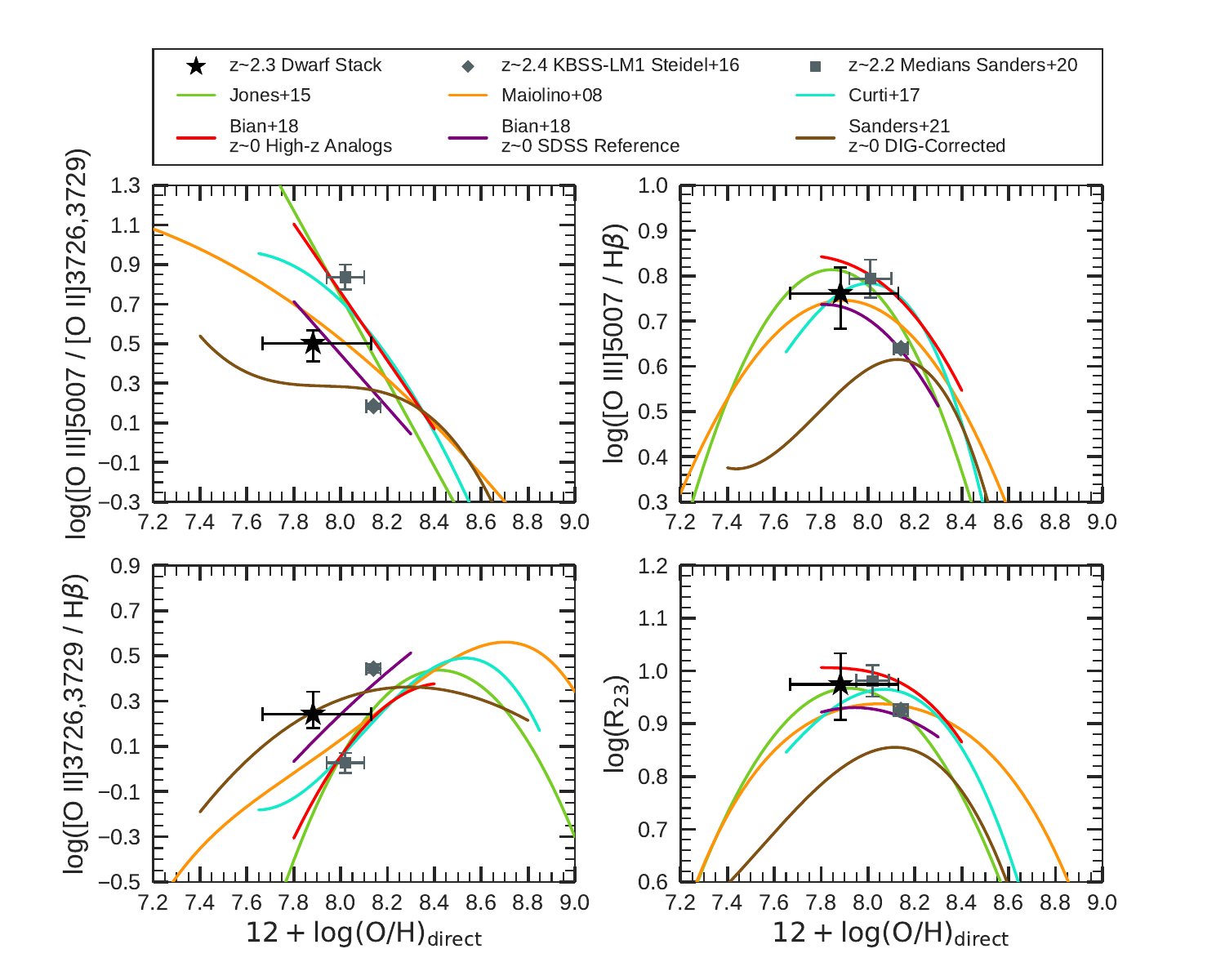}
    \caption{Strong emission-line ratios as a function of direct-method metallicity. From left-to-right and top-to-bottom we consider the oxygen-based strong-line ratios of $\rm{O_{32}}$, $\rm{O_3}$, $\rm{O_2}$ and $\rm{R_{23}}$. Our $z \sim 2.3$ composite is displayed via the black star in each panel. For reference, we also include the $z \sim 2.4$ KBSS-LM1 composite of 30 star-forming galaxies also selected independent of emission-line strength \citep[gray diamond;][]{Steidel2016}, as well as the median values of the $z \sim 2.2$ auroral-line sample of \citet[][gray square]{Sanders2020}. We compare these points to the locally-calibrated, strong-line metallicity relations of \citet[][orange line]{Maiolino2008}, \citet[][green line]{Jones2015}, \citet[][cyan line]{Curti2017}, \citet[][both the local reference$\,-\,$purple line$\,-\,$and high-$z$ analog$\,-\,$red line$\,-\,$relations]{Bian2018}, and \citet[][DIG-corrected; brown line]{Sanders2021}. We note the metallicity-insensitivity of the $\rm{O_3}$ and $\rm{R_{23}}$ indices at the oxygen abundances considered here. We find that our composite and KBSS-LM1 have their metallicities best reproduced by the local reference relations of Bian et al., in contrast to the auroral-line sample of Sanders et al. that favors the high-$z$ analog relations.  \newline \label{fig:slr_dir_metal}}
\end{figure*}

\subsection{Strong-Line Metallicity Calibrations at High-\textit{z}} \label{subsec:SLR_O/H}
A major outstanding issue and active area of research in high-$z$ astronomy is how to accurately calculate the gas-phase metallicities of the star-forming galaxies in the various large, statistical, spectroscopic surveys at $z > 1$ and $M_\ast \gtrsim 10^9\, \rm{M_\odot}$ (e.g., 3D-\textit{HST}; \citealt{Brammer2012_3dhst}, KBSS-MOSFIRE; \citealt{Steidel2014}, MOSDEF; \citealt{Kriek2015}, FMOS-COSMOS; \citealt{Kashino2019}). The cause of this problem is two-fold. For one, auroral lines such as \ion{O}{3}] $\lambda\lambda$1661, 1666 or [\ion{O}{3}] $\lambda$4363, needed for direct, $T_e$-based metallicity estimation, are exceedingly faint, especially at high-redshift and with increasing galaxy stellar mass. Additionally, it is not fully understood how changing physical conditions with redshift in star-forming regions affect locally-calibrated, strong-line ratio metallicity diagnostics for indirect metallicity estimation. In other words, the accuracy and applicability of these strong-line metallicity calibrations at high-redshift is an open question which several studies have tried to address \citep{Jones2015, Sanders2016o3, Patricio2018, Gburek2019, Sanders2020}.

In Figure \ref{fig:slr_dir_metal}, we revisit this issue with our $z \sim 2.3$ dwarf galaxy composite. Here we plot our stack (the black stars) in the parameter space of $T_e$-based oxygen abundance versus various commonly-used, oxygen-based, strong emission-line ratios. The dust-corrected emission-line ratios and direct metallicity ($12+\log(\rm{O/H}) = 7.88^{+0.25}_{-0.22}$) of our stack are measured (see Section \ref{sec:measurements}) from the composite spectrum in Figure \ref{fig:comp_spectrum}. Our stack in these plots is compared to several locally-calibrated strong-line metallicity diagnostics from the literature \citep{Maiolino2008, Jones2015, Curti2017, Bian2018, Sanders2021} in an attempt to shed light on which strong-line ratios are serviceable at the typical metallicity of our stacking sample as well as which calibrations most favorably reproduce our composite metallicity at fixed strong-line ratio. While we cannot comment on the shape or slope of the various calibrations, we can get a sense of the appropriate normalization of these relations when considering high-$z$ dwarf galaxies. In Figure \ref{fig:slr_dir_metal}, we consider four strong-line ratios: $\rm{O_{32}}$ = log([\ion{O}{3}] $\lambda$5007/[\ion{O}{2}] $\lambda\lambda$3726, 3729), $\rm{O_3}$ = log([\ion{O}{3}] $\lambda$5007/H$\beta$), $\rm{O_2}$ = log([\ion{O}{2}] $\lambda\lambda$3726, 3729/H$\beta$), and $\rm{R_{23}}$ = log(([\ion{O}{3}] $\lambda\lambda$4959, 5007 + [\ion{O}{2}] $\lambda\lambda$3726, 3729)/H$\beta$). We do not consider strong-line metallicity calibrations based on [\ion{N}{2}] in this work due to our low detection significance ($\sim 2\sigma$) of [\ion{N}{2}] $\lambda$6583 in the composite spectrum as well as concerns in the literature \citep[e.g.,][]{Masters2014, Masters2016} of elevated N/O abundance ratios at high-redshift.

Of immediate note when considering the location of our stack relative to the $\rm{O_3}$- and $\rm{R_{23}}$-based calibrations in the right-hand panels of Figure \ref{fig:slr_dir_metal} is that our stack lies at or near the apex of these relations in what is called the ``turnover" between the high and low metallicity branches of these calibrations. In these turnover regimes, the strong-line ratio is insensitive to the metallicity of a galaxy, giving these relations little value as useful metallicity indicators over the oxygen abundance range spanned by the turnover region (somewhere roughly between $7.7 \lesssim \rm{12+log(O/H)} \lesssim 8.3$ depending on the strong-line index$\,-\, \rm{O_3}$ or $\rm{R_{23}} \,-\,$ and calibration used). We therefore do not recommend the use of these strong-line indices for $z \sim 2$ dwarf galaxies similar to our those in our stacking sample. These results and conclusion are not particularly surprising as [\ion{O}{3}] $\lambda$4363-emitter studies \citep{Sanders2016o3, Gburek2019, Sanders2020} and photoionization modeling \citep[e.g.,][]{Steidel2014} have shown that it is quite common for $z \sim 2$ star-forming galaxies, over a couple orders of magnitude in stellar mass, to have metallicities that lie within these insensitive turnover regions.

A more interesting result is revealed when looking at the dwarf galaxy stack relative to the $\rm{O_{32}}$- and $\rm{O_2}$-based calibrations in the left-hand panels of Figure \ref{fig:slr_dir_metal}. In particular, we focus on the comparisons with the strong-line metallicity relations of \citet{Bian2018}, who used stacked SDSS spectra to create $T_e$-based, empirical metallicity calibrations over the metallicity range $7.8 < \rm{12+log(O/H)} < 8.4$. These calibrations were created from two distinct SDSS samples, a reference sample of galaxies lying within $\pm0.05$ dex of the $z \sim 0$ star-forming sequence of the N2-BPT diagram (parameterized by \citealt{Kewley2013}) and a high-redshift analog sample of SDSS galaxies lying within $\pm0.04$ dex of the offset $z \sim 2.3$ star-forming sequence of the BPT diagram \citep{Steidel2014}. With these selection criteria based on nebular emission-line ratios, the calibrations of \citet{Bian2018} should represent the conditions of star-forming regions in low- and high-redshift galaxies, respectively. We note that the Bian et al. calibrations presented in Figure \ref{fig:slr_dir_metal} were re-fit for this study for the reasons, and via the methods, described in Appendix \ref{sec:b18_appendix}.

When comparing our stack (black stars) to the strong-line metallicity relations of \citet{Bian2018}, we find that our stack favors the local reference calibrations (purple curves), such that when considering the $\rm{O_{32}}$ and $\rm{O_2}$ indices, the local reference relations reproduce our composite metallicity to within $\lesssim 0.12$ dex at fixed strong-line ratio. The high-$z$ analog relations of Bian et al. (red curves) and the calibrations of \citet[][orange curves]{Maiolino2008}, \citet[][green curves]{Jones2015}, and \citet[][cyan curves]{Curti2017} are all $\gtrsim 1\sigma$ inconsistent with our stack in the $\rm{O_{32}}$ and $\rm{O_2}$ panels, the lone exception being the $\rm{O_{32}}$ relation of Maiolino et al. (We remind the reader that the $1\sigma$ uncertainties of the composite metallicity and strong-line ratios include both statistical error and sample variance; see Section \ref{subsec:stack}.) Additionally, in all panels other than that of the $\rm{O_2}$ index, our composite is also highly inconsistent with the calibrations of \citet[][brown curves]{Sanders2021}, which, at metallicities below $\rm{12+log(O/H)} < 8.4$, are calibrated with the \ion{H}{2} region spectra of dwarf galaxies from \citet{Berg2012} and the \textit{Spitzer} Local Volume Legacy survey \citep{Dale2009}. These dwarf galaxies have been shown in previous studies to not follow the other strong-line metallicity relations considered in this work, possibly as a result of biases from the selection methods of the various calibration samples or an incompleteness in low-metallicity, high-excitation \ion{H}{2} regions \citep{Gburek2019, Sanders2020, Sanders2021}.

It is interesting that our $z \sim 2.3$ stack generally best agrees with the local reference calibrations of \citet{Bian2018}, particularly when compared to the findings of \citet{Sanders2020}, who compiled \ion{O}{3}] $\lambda\lambda$1661, 1666 and [\ion{O}{3}] $\lambda$4363-detected sources at $z > 1$ from the literature and the MOSDEF survey and conducted a similar study of strong-line metallicity diagnostics at the median redshift of the compiled sample, $z_{\rm{med}} \sim 2.2$. These authors found that at the median metallicities and median line ratios of the galaxies in their $z > 1$ sample (gray squares in Figure \ref{fig:slr_dir_metal}; each median limited to galaxies with detections of the respective line ratio), the high-redshift analog calibrations of \citet{Bian2018} best reproduced their metallicities at fixed line-ratio. Moreover, \citet{Sanders2020} found general agreement between their median points and the \citet{Curti2017} calibrations as well as general agreement with the calibration sample ($z \sim 0$ SDSS galaxies from \citealt{Izotov2006}) of the \citet{Jones2015} relations. Consequently, when considering the results of our work and those of \citet{Sanders2020}, both studies at similar redshift and metallicity, a tension exists in regard to the evolution with redshift of strong-line metallicity diagnostics and therefore which calibrations are reliable at high-$z$.

A likely source of the discrepancy and tension seen between the results of this work and that of \citet{Sanders2020} lies in how each galaxy sample was selected. For our dwarf galaxy stacking sample, we did not select galaxies based on the strength of any particular rest-optical emission-line (see Section \ref{subsec:samplesel}). This is important particularly when considering [\ion{O}{3}] $\lambda$4363, [\ion{O}{3}] $\lambda$5007 ([\ion{O}{3}] $\lambda$4959 in our case), and H$\alpha$. By avoiding selection based on line-strength, we mitigate biases in our sample such as high sSFRs (SFR$\,$/$\, M_\ast$) and high excitation, resulting in a sample which very nearly falls on an extrapolation of the mean $z \sim 2.3$ $M_\ast-\rm{SFR}$ relation (see Section \ref{subsubsec:SFMS} and Figure \ref{fig:SFMS}). In contrast, the $z \sim 2.2$ sample of \citet{Sanders2020} comprises galaxies selected on auroral-line detection (either \ion{O}{3}] $\lambda\lambda$1661, 1666 or [\ion{O}{3}] $\lambda$4363, detected at $\rm{(S/N)_{med}} = 6.0_{-2.5}^{+1.5}$) for $T_e$-based metallicity estimation. Due to the faintness of these auroral lines$\,-\,$[\ion{O}{3}] $\lambda$5007 is $\sim 30 - 100\times$ brighter than [\ion{O}{3}] $\lambda$4363 \citep{Jones2015}$\,-\,$this detection requirement preferentially selects younger galaxies with high-excitation, highly-ionized star-forming regions. These galaxies lie well above the $M_\ast-\rm{SFR}$ relation. In fact, the sample of \citet{Sanders2020} lies an average $\sim 0.6$ dex above the best-fit $z \sim 2.3$ $M_\ast-\rm{SFR}$ relation of \citet{Sanders2018} (which is very similar to the fit of the same relation in \citealt{Sanders2021}) and has $\rm{O_{32}}$ values $\sim 0.5$ dex higher on average than typical $z \sim 2.3$ star-forming MOSDEF galaxies at fixed $M_\ast$ \citep[][Figure 16]{Sanders2020}. Sanders et al. shows that this auroral-line sample is not representative of typical $z \sim 2.3$ star-forming galaxies, but rather coincident with $z \sim 2$ extreme emission-line galaxies (EELGs). In effect, at roughly fixed O/H in Figure \ref{fig:slr_dir_metal}, we are comparing two galaxy samples that differ significantly in the ionization state and extremity of their star-forming regions. As such, both samples are not well-represented by one single strong-line metallicity calibration.

When considering the sample selection methodologies for our stacked sample and the $z > 1$ auroral-line sample of \citet{Sanders2020} against those for the calibration samples used to parameterize the relations shown in Figure \ref{fig:slr_dir_metal}, it is perhaps not surprising that several relations appear to agree with the Sanders et al. auroral-line sample while being $\gtrsim 1\sigma$ inconsistent with our stacked sample. In the low-metallicity regime ($12+\log\rm{(O/H)} \lesssim 8.4$), the locally-calibrated strong-line metallicity relations are generally defined by \textit{individual} galaxies with [\ion{O}{3}] $\lambda$4363 detections. This is the case for the relations of \citet{Maiolino2008}, \citet{Jones2015}, and \citet{Curti2017}. This requirement of an [\ion{O}{3}] $\lambda$4363 detection in individual galaxies, as well as the BPT-related requirement of the \citet{Bian2018} high-$z$ analog calibration (detailed above), selectively probe galaxies with ISM conditions more extreme than in typical $z \sim 0$ star-forming galaxies. Rather, ISM conditions more akin to those in the $z > 1$ auroral-line sample are probed, leading to low-metallicity strong-line calibrations that are likely biased high in $\rm{O_3}$ and $\rm{O_{32}}$ (and low in $\rm{O_2}$) and that closely predict the \citet{Sanders2020} $z > 1$ sample metallicity at fixed strong-line ratio. Meanwhile, the selection criterion for the \citet{Bian2018} local reference calibrations (detailed above) selectively probes galaxies with less extreme star-forming conditions and better predicts the metallicity at fixed strong-line ratio of our dwarf galaxy stack, which is not reliant on emission-line detections.\footnote{A more detailed study of the biases in certain locally-calibrated strong-line metallicity relations (particularly of \citealt{Curti2017}), and how those biases factor into the observed evolution of strong-line metallicity relations with redshift, can be found in \citet{Sanders2020}.}

\subsubsection{The KBSS-LM1 Composite of \citet{Steidel2016}} \label{subsubsec:kbss-lm1}
In this subsection, we briefly discuss the $z \sim 2.4$ composite spectrum of \citet{Steidel2016} in the context of the strong-line ratio$\,-\,$direct metallicity relations displayed in Figure \ref{fig:slr_dir_metal}. This composite, referred to in \citet{Steidel2016} as ``KBSS-LM1," is derived from the rest-frame far-UV and optical spectra of 30 star-forming galaxies from the KBSS-MOSFIRE spectroscopic survey. The galaxies comprising KBSS-LM1 were notably selected to have emission-line measurements or limits (not detections) of, among other lines, [\ion{O}{2}] $\lambda\lambda$3726, 3729, H$\beta$, [\ion{O}{3}] $\lambda\lambda$4959, 5007, and H$\alpha$. The sample of galaxies was also selected such that it broadly represents the full KBSS-MOSFIRE sample in SFR, $M_\ast$, and O/H, the latter of which was calculated via the $T_e$-sensitive UV emission-line doublet, \ion{O}{3}] $\lambda\lambda$1661, 1666. The median $M_\ast$ and derived $T_e$-based metallicity of KBSS-LM1 are, as reported by \citet{Sanders2020}, log($M_\ast/\rm{M_\odot}$) = 9.8 $\pm$ 0.3 and 12+log(O/H) = 8.14 $\pm$ 0.03, respectively. With these selection criteria, KBSS-LM1 is therefore similar to our $z \sim 2.3$ dwarf galaxy composite in the sense that both samples should be representative of typical $z \sim 2.3$ galaxies at their respective stellar masses. Indeed, \citet[][Figure 16]{Sanders2020} find KBSS-LM1 to lie just above the $z \sim 2.3$ star-forming main sequence as well as amongst the typical $z \sim 2.3$ MOSDEF galaxies in $\rm{O_{32}}$ at fixed $M_\ast$.

In Figure \ref{fig:slr_dir_metal}, along with our $z \sim 2.3$ dwarf galaxy stack (black stars) and $z \sim 2.2$ auroral-line sample of \citet[][gray squares]{Sanders2020}, we plot KBSS-LM1 as a gray diamond in each panel. As with our dwarf galaxy stack, albeit at higher O/H owing to its higher $M_\ast$, we see that KBSS-LM1 is best represented by the local reference calibrations of \citet{Bian2018} instead of the high-redshift analog calibrations, though note the small statistical error bars (not bootstrapped) of KBSS-LM1. This result suggests that $z \sim 2.3$ star-forming galaxies with $8.4 \lesssim \log(M_\ast/\rm{M_\odot}) \lesssim 9.8$ that lie on the $M_\ast$-SFR relation will, on average, have their metallicities most accurately predicted at fixed strong-line ratio via the local reference strong-line metallicity calibrations of \citet{Bian2018}. In contrast, $z \sim 2.3$ EELGs and galaxies with more extreme ISM conditions, like the $z \sim 2.2$ auroral-line sample, may require independently-calibrated strong-line metallicity relations more akin to the high-$z$ analog calibrations of \citet{Bian2018}.

\begin{figure*}[ht!]
    \includegraphics[trim={0cm 0.2cm 0.2cm 0.2cm}, width=1.05\columnwidth, clip]{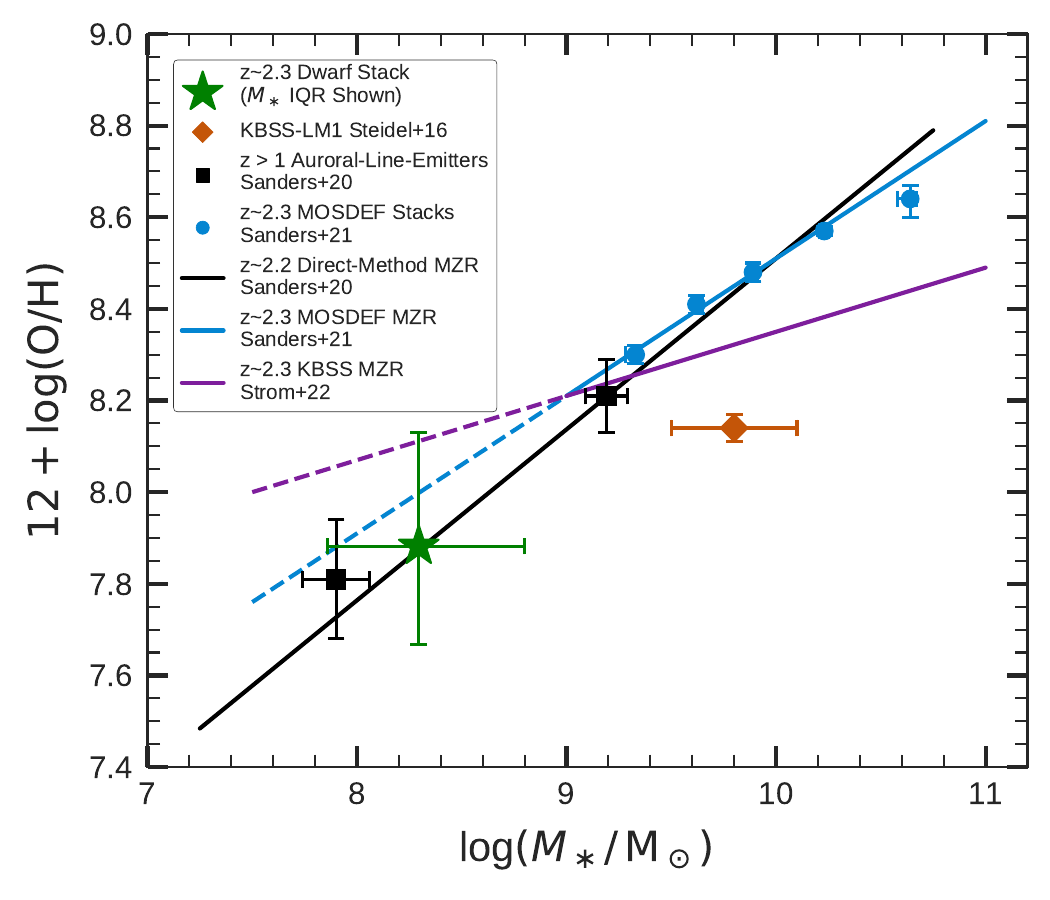}\quad\includegraphics[trim={0cm 0.2cm 0.2cm 0.2cm}, width=1.05\columnwidth, clip]{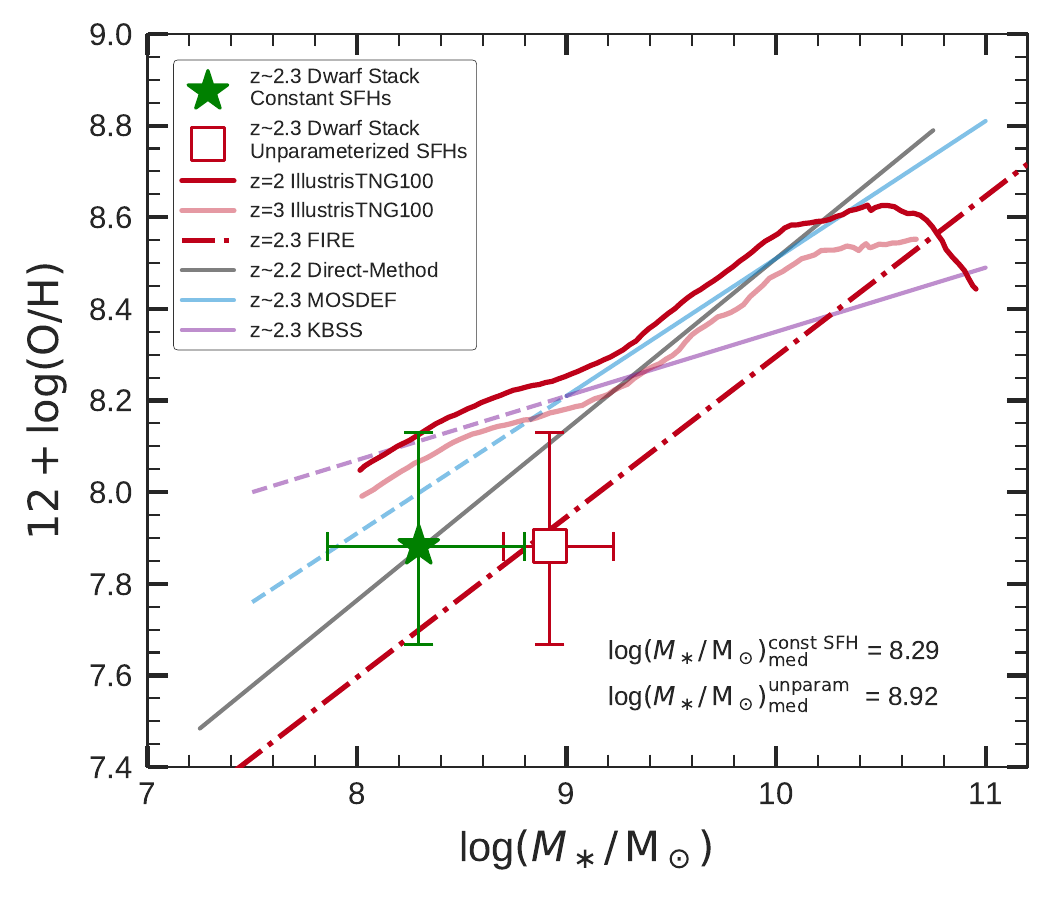}
    \caption{The stellar mass$\,-\,$gas-phase metallicity relation (MZR). \textit{Left}: We compare our $z \sim 2.3$ dwarf galaxy stack (green star) at our sample's median mass, log($M_\ast$/$\rm{M_\odot}$)$_{\rm{med}} = 8.29^{+0.51}_{-0.43}$, against the $z \sim 2.2$ direct-method MZR of \citet[][black line]{Sanders2020} and the extrapolations of the $z \sim 2.3$ strong-line MZR of \citet[][blue solid/dashed line]{Sanders2021} and the $z \sim 2.3$ photoionization-model-based MZR of \citet[][purple solid/dashed line]{Strom2022}. We find excellent agreement between our stack and the direct-method MZR, the most comparable of the relations. For reference, we also include the mean values of the $z > 1$ auroral-line sample split into two $M_\ast$ bins \citep[black squares;][]{Sanders2020}, the MOSDEF galaxy stacks from \citet[][blue squares]{Sanders2021}, and the $z \sim 2.4$ KBSS-LM1 composite of \citet[][orange diamond]{Steidel2016}. \textit{Right}: We compare our $z \sim 2.3$ stack (green star; same as in left panel) to the predicted $z=2$ and $z=3$ MZRs (dark-red and light-red lines, respectively) of the IllustrisTNG100 simulations \citep{Torrey2019} and the $z=2.3$ MZR of the FIRE simulations \citep[red dot-dashed line;][]{Ma2016}, finding consistency, within errors, with both sets of MZRs. When comparing our stack after recalculating the sample stellar masses assuming more realistic non-parametric SFHs (red unfilled square; log($M_\ast$/$\rm{M_\odot}$)$_{\rm{med}}$ = $8.92^{+0.31}_{-0.22}$), we find excellent agreement with the MZR from FIRE. The MZRs from the left-hand panel are recreated here for reference. We note that for each display of our stack, the $x$-axis error bar corresponds to the $M_\ast$ interquartile range (IQR) of our sample. \newline \label{fig:MZRs}}
\end{figure*}

\subsection{The Stellar Mass$\,-\,$Gas-Phase Metallicity Relation} \label{subsec:z2.3_mzr}
This section explores our $z \sim 2.3$ dwarf galaxy sample in relation to the low-mass end of the stellar mass$\,-\,$gas-phase metallicity relation (MZR). The MZR, shown to exist both locally and at high redshift, displays a positive correlation, parameterized as a power-law, between galaxy stellar mass ($M_\ast$) and gas-phase oxygen abundance (O/H) at lower stellar masses before flattening asymptotically at higher masses \citep[$M_\ast \gtrsim 10^{10.5}\ \rm{M_\odot}$ locally;][]{Tremonti2004, Erb2006, Maiolino2008, Andrews&Martini2013, Steidel2014, Curti2020, Sanders2020, Sanders2021, Strom2022}. The MZR also evolves with time, such that at higher redshifts, the average metallicity of star-forming galaxies is lower at fixed $M_\ast$ than it is locally \citep{Erb2006, Maiolino2008, Zahid2013, Zahid2014_aug, Zahid2014_sept, Steidel2014, Sanders2020, Sanders2021}. Here we compare our dwarf galaxy composite against recent empirical and theoretical parameterizations of the MZR in an effort to better constrain the low-mass slope and normalization of the relation at $z \sim 2.3$. 

\subsubsection{Comparison to MZRs Derived from Observations} \label{subsubsec:emp_mzrs}
In the left-hand panel of Figure \ref{fig:MZRs}, we plot our $z \sim 2.3$ dwarf galaxy composite against the $z \sim 2.2$ direct-method MZR of \citet{Sanders2020}, the $z \sim 2.3$ strong-line MZR of \citet{Sanders2021}, and the $z \sim 2.3$ MZR of \citet{Strom2022}, who calculated their metallicities via photoionization modelling. The direct-method MZR shown here (black line) is derived from the \citet{Sanders2020} $z > 1$ ($z_{\rm{med}} \sim 2.2$) auroral-line sample considered in Section \ref{subsec:SLR_O/H} above. However, as previously mentioned, this sample of galaxies lies $\langle\Delta\log(\rm{SFR})\rangle\sim0.6$ dex above the $z \sim 2.3$ $M_\ast\,-\,$SFR relation, which has the effect, via the FMR, of yielding a MZR biased low in O/H relative to the MZR expected for typical galaxies that fall on the star-forming main sequence at this redshift. As such, the direct-method MZR displayed in Figure \ref{fig:MZRs} was parameterized by \citet{Sanders2020} \textit{after} applying SFR-corrections to the $z > 1$ sample metallicities via Equation \ref{equ:o/h_sfr_relation}. Additionally, \citet{Sanders2020} adjusted this MZR to account for a low-redshift bias in the low-mass $z > 1$ galaxies. These galaxies (low-mass black square; 5 galaxies) were found to have a median redshift $\Delta z_{\rm{med}} \sim 0.5$ lower than the high-mass sample galaxies (high-mass black square; 9 galaxies), and were estimated, on average, to have a metallicity biased 0.1 dex high due to the redshift evolution of O/H at fixed $M_\ast$ \citep[$d$log(O/H)/$dz \approx -0.2$;][]{Sanders2020}. With these adjustments, the direct-method MZR reproduced in Figure \ref{fig:MZRs} is an estimation of the MZR$\,-\,$on the $T_e$-based abundance scale$\,-\,$of typical galaxies lying on the star-forming main sequence at a redshift of $z \sim 2.2$.

The blue line (and its extrapolation) in the left-hand panel of Figure \ref{fig:MZRs} is a recreation of the $z \sim 2.3$ strong-line MZR of \citet{Sanders2021}, fit as a power-law to the four lowest-mass bins of MOSDEF galaxies displayed here as blue squares (The highest-mass bin suffers from incompleteness and was excluded from the fit \citep[][Section 2.4]{Sanders2021}). These MOSDEF stacks were also used by Sanders et al. to parameterize their $M_\ast-\rm{SFR}$ relation (see Section \ref{subsubsec:SFMS} and Figure \ref{fig:SFMS}). The metallicities of the MOSDEF stacks were calculated via the high-$z$ analog calibrations of \citet{Bian2018} and the $\alpha$-element-based strong-line ratios of $\rm{O_{32}}$, $\rm{O_3}$, and log([\ion{Ne}{3}] $\lambda$3869/[\ion{O}{2}] $\lambda\lambda$3726, 3729). These relations were chosen so as to use calibrations that most closely reproduce the excitation sequences, and thus likely ISM conditions, of the MOSDEF $z \sim 2.3$ star-forming sample.

The $z \sim 2.3$ MZR of \citet{Strom2022}, and its extrapolation, are shown as the purple line in the left-hand panel of Figure \ref{fig:MZRs}. This relation was fit to 195 individual star-forming galaxies from the KBSS survey. The metallicities in this study were estimated with photoionization models, described in \citet{Strom2018, Strom2022}, that are able to reproduce the rest-UV and rest-optical spectroscopic properties of high-$z$, star-forming galaxies. 


When comparing our $z \sim 2.3$ dwarf galaxy stack (green star; $x$-axis error bar represents the $M_\ast$ interquartile range of the stacking sample) against the direct-method and strong-line MZRs of \citet{Sanders2020} and \citet{Sanders2021}, respectively, we find that at the median stellar mass of our stacking sample, log($M_\ast$/$\rm{M_\odot}$)$_{\rm{med}} = 8.29^{+0.51}_{-0.43}$, and the direct-method metallicity calculated from the composite spectrum, $12+\log(\rm{O/H}) = 7.88^{+0.25}_{-0.22}$, the stack lies virtually on top of the $z \sim 2.2$ direct-method MZR. Within uncertainties, our stack \citep[and the mean values of the low- and high-mass bins of the $z > 1$ auroral-line sample;][black squares]{Sanders2020} is also consistent with the extrapolation of the strong-line MZR, lying below this extrapolation by $\sim 0.12$ dex. Correcting the metallicity of our composite for the slightly-high SFR bias of the stacking sample (increasing log(O/H) by $\sim 0.06$ dex; see Section \ref{subsubsec:SFMS}) reduces the offset of the stack from the extrapolation of the strong-line MZR while moving the stack $\sim 0.07$ dex above the direct-method MZR. This consistency with both MZRs supports their fit power-law parameters, in particular the slope, which was fit by \citet{Sanders2020} to be $\beta = 0.37$ and by \citet{Sanders2021} to be $\beta = 0.30$. However, while we are comparing to the strong-line MZR of \citet{Sanders2021} due to the authors' careful consideration in selecting applicable strong-line calibrations for high-$z$ star-forming regions, we note that systematic uncertainties still exist between metallicities calculated directly versus with strong-line proxies. Therefore, in constraining the slope of the $z \sim 2.3$ MZR, our results and direct-method of metallicity estimation most favorably suggest the slope fit to the direct-method MZR of \citet[][]{Sanders2020}, $\beta = 0.37$.

When considering the $z \sim 2.3$ photoionization-model-based MZR of \citet{Strom2022}, we find that our stack lies $\approx 1\sigma$ below the extrapolation of this relation, in disagreement with the shallow slope ($\beta = 0.14$) proposed by Strom et al. However, we must again take into account the difference in methods of abundance estimation between the two studies. As discussed in the literature \citep[e.g.,][]{Kewley&Ellison2008, Maiolino&Mannucci2019}, photoionization models typically overestimate metallicities by $\sim 0.2$ dex or more compared to direct metallicities. In general, this is due to a poorly-constrained combination of factors, such as photoionization models accounting for dust depletion or $T_e$-based metallicities potentially being biased low due to temperature fluctuations or gradients in star-forming regions leading to nebular spectra dominated by brighter auroral lines from high-$T_e$ zones. While the discrepancy between direct and theoretical metallicity estimates is stronger at higher metallicities \citep[e.g.,][]{Kewley&Ellison2008, Maiolino&Mannucci2019, Curti2020}, \citet{Steidel2016} found that photoionization models predict an oxygen abundance 0.25 dex higher than their $T_e$-based estimate of 12+log(O/H) = 8.14 (0.29 $\rm{Z_\odot}$) for KBSS-LM1, a composite of 30 star-forming KBSS galaxies that we briefly discussed in Section \ref{subsubsec:kbss-lm1}. If we apply this same offset to the $T_e$-based metallicity of our dwarf galaxy composite, our stack will lie very near the extrapolation of the \citet{Strom2022} MZR. This said, there are too many systematic uncertainties involved to accurately compare our composite against this MZR at present, and we conclude that the best estimate for the slope of the $z \sim 2.3$ MZR is $\beta = 0.37$, given by the direct-method MZR of \citet{Sanders2020}.

We note that the stellar masses of the samples and studies considered here are calculated via SED-fitting with consistent assumptions in star-formation history, IMF, and extinction law. We also note that the KBSS-LM1 composite of \citet{Steidel2016} lies significantly below the displayed MZRs, including the direct-method MZR, when plotted at its reported direct metallicity of 12+log(O/H) = $8.14 \pm 0.03$. However, we do acknowledge the very small \textit{statistical-only} uncertainty of this metallicity.

\subsubsection{Comparison to MZRs from Cosmological Simulations} \label{subsubsec:sim_mzrs}
In the right-hand panel of Figure \ref{fig:MZRs}, we now compare our $z \sim 2.3$ dwarf galaxy stack against predicted MZRs from well-known cosmological simulations, the Feedback in Realistic Environments simulations\footnote{FIRE: \url{https://fire.northwestern.edu/}} \citep[FIRE;][]{Hopkins2014} and The Next Generation Illustris simulations\footnote{IllustrisTNG: \url{https://www.tng-project.org/}} \citep[IllustrisTNG;][]{Weinberger2017, Pillepich2018}, which are the successor to the Illustris simulation \citep{Vogelsberger2014, Vogelsberger2014_nature, Genel2014}. Our stack, with the median mass (log($M_\ast$/$\rm{M_\odot}$) = $8.29^{+0.51}_{-0.43}$) and direct metallicity ($12+\log(\rm{O/H}) = 7.88^{+0.25}_{-0.22}$) as reported throughout this paper, is shown by the green star, with the $x$-axis error bar representing the $M_\ast$ interquartile range of our stacking sample. The predicted MZR from FIRE, given by the redshift-dependent, fixed-slope ($\beta=0.35$), gas-phase MZR fitting function in \citet{Ma2016}, is evaluated at $z=2.30$ (the mean redshift of our stacking sample) and is represented by the red dot-dashed line. The $z=2$ and $z=3$ MZRs from IllustrisTNG \citep{Torrey2019}, specifically from the TNG100 simulation volume, are displayed by the solid dark-red and light-red lines, respectively. For reference to MZRs derived from observations, in the right-hand panel we reproduce, from the left-hand panel, the $z \sim 2.2$ direct-method MZR of \citet[][black line]{Sanders2020}, the $z \sim 2.3$ strong-line MZR of \citet[][blue line]{Sanders2021}, and the $z \sim 2.3$ photoionization-model-based MZR of \citet[][purple line]{Strom2022}. 

Properly assessing the predicted MZRs considered in this section requires a reliable and accurate empirical metallicity estimation method such as the direct-method, which estimates oxygen abundances through directly probing physical properties ($T_e$ and $n_e$) of star-forming regions. We find that at fixed $M_\ast$, the direct metallicity of our stack (green star) is consistent within uncertainties with both the $z=2.3$ FIRE MZR and $2 \leqslant z \leqslant 3$ MZR of IllustrisTNG. We also observe that the $z \sim 2.2$ direct-method MZR agrees favorably in slope and normalization with the IllustrisTNG MZR above $M_\ast \approx 10^9\ \rm{M_\odot}$. Below this mass, the IllustrisTNG MZR deviates away from this slope and the direct-method MZR toward higher metallicity values, displaying a ``bump" in the low-mass regime. This bump is the result of the minimum wind velocity ($v_{\rm{min}}=350\ \rm{km}\ \rm{s^{-1}}$) enforced in the stellar feedback models of IllustrisTNG, with $v_{\rm{min}}$ put in place so that the simulations match the low-end of the galaxy stellar mass function \citep{Pillepich2018}. As \citet{Torrey2019} explain, while $v_{\rm{min}}$ is not directly a function of $M_\ast$, it is generally set as the wind velocity in galaxies with $M_\ast\lesssim10^9\ \rm{M_\odot}$ due to the low dark matter velocity dispersions in their halos. The higher metallicities at these masses that we see as the bump in the MZR then arise because these fixed-velocity winds eject less gas, and therefore fewer metals, than would be the case if the wind velocity were allowed to be $v < 350\ \rm{km}\ \rm{s^{-1}}$. While further observations are needed to either confirm or deny this bump in the MZR, as well as evaluate the applicability of the minimum wind velocity assumption, the direct-method MZR of \citet{Sanders2020} suggests that such a bump likely does not exist and that wind velocities in low-mass galaxies can extend lower than $v < 350\ \rm{km}\ \rm{s^{-1}}$. This suggestion is reinforced below when revisiting our dwarf galaxy stacking sample with stellar masses recalculated under more realistic assumptions.

In this work, in order to facilitate fair comparisons of our empirical results to those in the literature, when estimating the stellar masses of our stacking sample via SED-fitting, we made assumptions consistent with those generally found in the literature; in particular, we assumed constant star formation histories (SFH). However, the SFHs of galaxies, particularly of galaxies at high-$z$ owing to their higher-EW emission lines, are likely not well-described by such simple parameterizations. Instead, stellar masses are likely more accurate if calculated assuming non-parametric SFHs which can better reveal the presence of older stellar populations that are hidden in the rest-UV and rest-optical by brighter, younger stars \citep{Gburek2019, Tang2022, Whitler2022, Topping2022}. Having these more realistic, typically larger stellar masses is important when comparing to simulation results. Therefore, we recalculated the stellar masses of our stacking sample, via the \texttt{PROSPECTOR}\footnote{\url{https://prospect.readthedocs.io/en/latest/}}$^{,}$\footnote{\url{https://github.com/bd-j/prospector}} SED-fitting code \citep{Johnson2021}, assuming non-parametric SFHs (see Table \ref{tab:stacking_sample}). The resultant median stellar mass and stellar mass interquartile range of our sample under these assumptions becomes log($M_\ast$/$\rm{M_\odot}$) = $8.92^{+0.31}_{-0.22}$, an increase to our fiducial median stellar mass of 0.63 dex. Our dwarf galaxy stack, shifted to this recalculated median $M_\ast$, is shown in Figure \ref{fig:MZRs} as the red unfilled square. As with the $x$-axis error bar of our fiducial point (the green star), the $x$-axis error bar of the red square represents the recalculated interquartile range of $M_\ast$ in our stacking sample.

When comparing our $z \sim 2.3$ dwarf galaxy stack to the simulated MZRs following recalculating the sample stellar masses under the assumption of more realistic non-parametric SFHs, we find excellent agreement between our composite and the $z=2.3$ FIRE MZR of \citet{Ma2016}, especially if we apply the metallicity correction ($\Delta \log(\rm{O/H}) \approx 0.06$ dex) to our stack to account for our bias high in SFR. In comparison to the $z=2$ and $z=3$ IllustrisTNG MZRs of \citet{Torrey2019}, we find that our composite lies $\sim 1.6\sigma$ below the $z=2$ MZR and $\sim 1.3\sigma$ below the $z=3$ MZR, suggesting that a stronger O/H evolution at fixed $M_\ast$ with redshift (like that seen in Figure 7 of \citealt{Ma2016} relative to other simulations of the time$\,-\,$including the original Illustris simulation \citealt{Torrey2014}) and/or tuning of the $z=0$ MZR normalization is needed. Like the direct-method MZR of \citet{Sanders2020}, our composite also disagrees with the existence of a metallicity bump in the MZR at $M_\ast\lesssim10^9\ \rm{M_\odot}$, suggesting that the minimum wind velocity assumption in IllustrisTNG should be revisited.

\subsection{$M_\ast-\rm{SFR}-\rm{O/H}$ Relation at $z \sim 2.3$} \label{subsec:FMR_z23}
The $M_\ast-\rm{SFR}-\rm{O/H}$ relation, or fundamental metallicity relation (FMR), posits that the MZR has a secondary dependence on SFR such that, when considering all three properties, the scatter in metallicity at fixed $M_\ast$ and SFR is reduced compared to the scatter in metallicity at fixed $M_\ast$ alone \citep{Mannucci2010, Lara-Lopez2010, Andrews&Martini2013, Sanders2018, Curti2020, Sanders2021}. Also as a result of this secondary dependence, at fixed $M_\ast$, a higher (lower) than average SFR yields a lower (higher) than average O/H. Of further interest is that numerous studies have shown that the FMR is redshift-invariant to within $\sim 0.1$ dex in metallicity out to $z \sim 3.3$ \citep[e.g.,][]{Mannucci2010, Henry2013_june, Henry2013_oct, Sanders2018, Sanders2021}. In effect, the observation of the evolution of the MZR over this redshift range, whereby O/H decreases with increasing redshift at fixed $M_\ast$, is actually the observation at different redshifts of different portions of the locally-defined FMR since SFR increases with redshift at fixed $M_\ast$ \citep{Whitaker2014, Speagle2014, Sanders2021}. Unfortunately, however, this redshift-invariance of the FMR is still a matter of debate owing to the uncertainties regarding the applicability of locally-calibrated, strong-line metallicity calibrations at high-redshift (see Section \ref{subsec:SLR_O/H}), widely-used in lieu of hard-to-measure direct metallicities.

Fortunately, in this work we have a direct, $T_e$-based oxygen abundance from our dwarf galaxy composite spectrum with which we can evaluate the redshift evolution of the FMR of low-mass galaxies. In doing so, we probe $M_\ast-\rm{SFR}-\rm{O/H}$ parameter space that to-date has been poorly-sampled at $z \sim 2.3$. We probe this space at high-$z$ via the commonly-used 2D planar projection of the 3D FMR, with a functional form first established by \citet{Mannucci2010}. In this projection, the metallicity is a function of the linear combination of $M_\ast$ and SFR, denoted by $\mu_\alpha$ and described by the equation

\begin{equation} \label{equ:fmr_projection}
    \mu_\alpha = \log(M_\ast/\mathrm{M_\odot}) - \alpha\log(\mathrm{SFR}/\mathrm{M_\odot\, yr^{-1}})
\end{equation}

\noindent where $\alpha$ is the parameter which denotes the strength of the SFR-dependence of the FMR as well as the value which minimizes the scatter in O/H at fixed $\mu_\alpha$. This parameter, $\alpha$, is generally found to be lower (a weaker SFR dependence) when determined with strong-line metallicities \citep[e.g., $\alpha=0.32$;][]{Mannucci2010} and higher (a stronger SFR dependence) when determined with direct-method metallicities \citep[e.g., $\alpha=0.66$;][]{Andrews&Martini2013}, though recent estimations via strong-line metallicities by \citet[][$\alpha=0.55$]{Curti2020} and \citet[][$\alpha=0.60$]{Sanders2021} have brought these $\alpha$-estimates into better agreement. For this work, we use the value of $\alpha=0.63$, which derives from the SDSS $M_\ast-\rm{SFR}$ stacks of \citet{Andrews&Martini2013}, with their direct-method metallicities corrected for diffuse ionized gas (DIG) contamination by \citet{Sanders2017}.

\begin{figure}[t]
    \includegraphics[trim={0.3cm 0.5cm 0.1cm 0cm}, width=\columnwidth, clip]{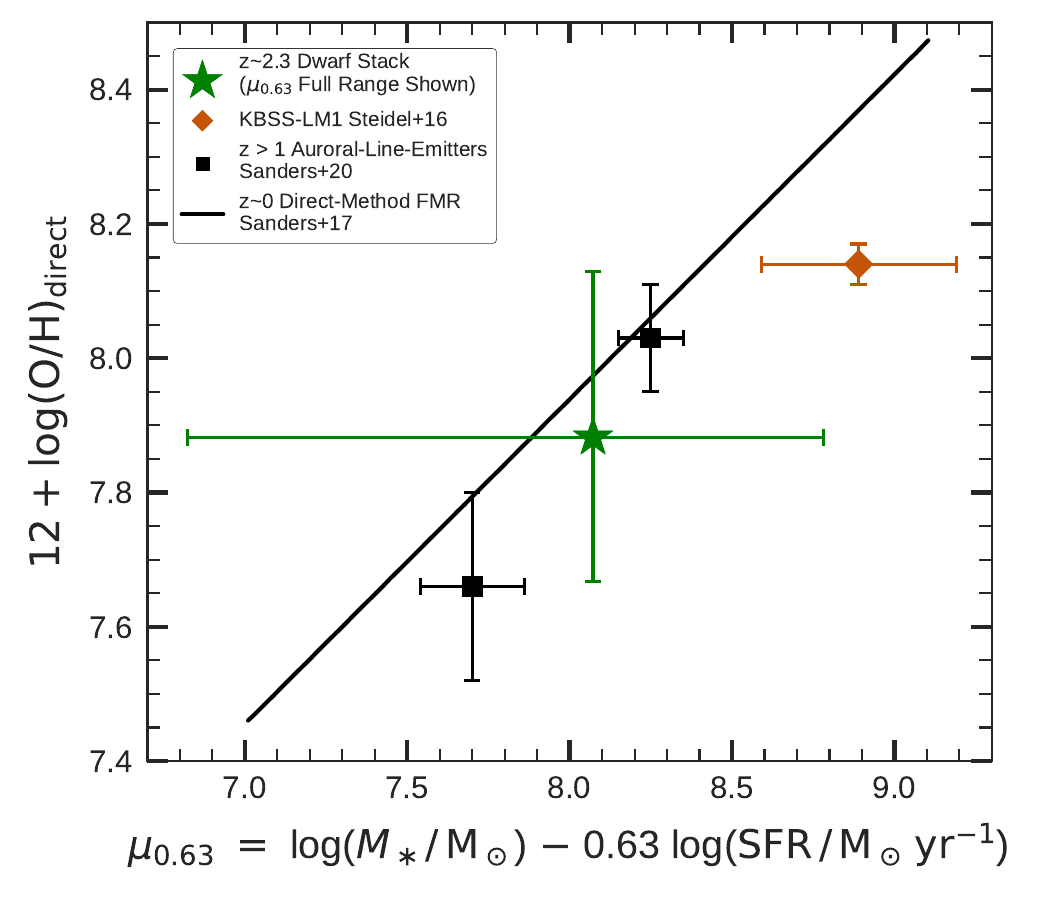}
    \caption{The direct-method $M_\ast-\rm{SFR}-\rm{O/H}$ fundamental metallicity relation (FMR), represented here by the O/H$\,-\,\mu_\alpha$ planar projection \citep{Mannucci2010}. Referencing Equation \ref{equ:fmr_projection} for $\mu_\alpha$, we take $\alpha=0.63$, the value determined by \citet{Sanders2017} using the DIG-corrected, $T_e$-based metallicities of the \citet{Andrews&Martini2013} SDSS $M_\ast-\rm{SFR}$ stacks. The best-fit linear relation by \citet{Sanders2017} to these $z \sim 0$ stacks is shown here by the black line. Our $z \sim 2.3$ composite is displayed as the green star with $\mu_{0.63}$ calculated from our sample's median $M_\ast$ (log($M_\ast$/$\rm{M_\odot}$) = 8.29) and SFR (log(SFR) = 0.353). The $x$-axis error bar represents our sample's full range in $\mu_{0.63}$. Our stack lies $\sim 0.10$ dex or $\sim 0.4\sigma$ below the direct-method FMR, though is consistent with a redshift-invariant FMR within uncertainties. For reference, we also show the $M_\ast$-binned, $z > 1$ auroral-line sample of \citet[][black squares]{Sanders2020} and the $z \sim 2.4$ KBSS-LM1 composite of \citet[][orange diamond]{Steidel2016}. \newline \label{fig:direct_FMR}}
\end{figure}

In Figure \ref{fig:direct_FMR}, we show our $z \sim 2.3$ dwarf galaxy stack (green star; $x$-axis error bar represents the full $\mu_{0.63}$ range of the stacking sample, [6.82, 8.78]) in the direct-method $\rm{O/H}-\mu_{0.63}$ parameter space, plotted at the $\mu_{0.63}$ value, $\mu_{0.63}=8.07$, given by the stacking sample's median stellar mass, log($M_\ast$/$\rm{M_\odot}$) = 8.29, and median SFR, log(SFR) = 0.353 (The median $\mu_{0.63}$ value of the individual galaxies comprising our stacking sample is similar at $\mu^{\rm{med}}_{0.63} = 8.11$). We compare our composite to the \citet{Sanders2017} best-fit linear representation of the FMR (black line) tracing the $z \sim 0$, DIG-corrected, \citet{Andrews&Martini2013} $M_\ast-\rm{SFR}$ stacks. We also include in Figure \ref{fig:direct_FMR} the two mass bins of the \citet{Sanders2020} $z > 1$ auroral-line sample (black squares), which these authors found to be consistent, within $\sim 0.1$ dex at fixed $\mu_{0.63}$, with lower-redshift samples in the $\mu_{0.63}$ direct-method FMR projection. Adding to these results, we find our dwarf galaxy composite to have a $T_e$-based metallicity that lies $\sim 0.4\sigma$ or $\sim 0.10$ dex below the linear relation representing the direct-method FMR. We therefore join numerous other authors in suggesting, via direct-method metallicities, that the FMR evolves at most by $\sim 0.1$ dex in O/H at fixed $M_\ast$ and SFR from $z=0$ to at least $z \sim 2.3$, though note that within the uncertainties on the metallicity, our stack is consistent with a redshift-invariant FMR.

\section{Summary} \label{sec:summary_ch3}
In this study, we analyze the median composite spectrum of 16 typical, star-forming, dwarf galaxies ($7.06 \leqslant \log(M_\ast/\rm{M_\odot}) \leqslant 8.93$; $\log(M_\ast/\rm{M_\odot})_{\rm{median}} = 8.29^{+0.51}_{-0.43}$) at redshifts $1.7 < z < 2.6$ ($z_{\rm{mean}} = 2.30$) selected independent of the strength of any particular emission line. These galaxies are gravitationally-lensed by the foreground clusters Abell 1689, MACS J0717.5+3745, and MACS J1149.5+2223. In our composite spectrum, we find a $2.5\sigma$ ($4.1\sigma$) detection of the faint, $T_e$-sensitive, [\ion{O}{3}] $\lambda$4363 auroral line when considering our bootstrapped (statistical-only) error spectrum, allowing us to directly calculate an oxygen abundance from the composite of $12+\log(\rm{O/H})_{\rm{direct}} = 7.88^{+0.25}_{-0.22}$ ($0.15^{+0.12}_{-0.06}\ \rm{Z_\odot}$). We summarize the results using this $T_e$-based metallicity, and other conclusions, in this final section. 

\begin{enumerate}
    \item To determine how representative our dwarf galaxy sample is of typical, star-forming, $z \sim 2.3$ dwarf galaxies, we first considered our composite in the context of the N2-BPT diagram, where we found that our stack lies offset from the $z \sim 0$ SDSS star-forming sequence in the same parameter space as the $z \sim 2.3$ star-forming galaxies of larger statistical surveys \citep[e.g., MOSDEF;][]{Shapley2015}. We also show that our composite lies at higher [\ion{O}{3}] $\lambda$5007/H$\beta$ and lower [\ion{N}{2}] $\lambda$6583/H$\alpha$ than any of the $M_\ast$-binned MOSDEF stacks of \citet{Sanders2021}; our composite extends the trend seen with these MOSDEF stacks of lower stellar mass and metallicity at higher [\ion{O}{3}]/H$\beta$ and lower [\ion{N}{2}]/H$\alpha$. 
    
    \item We also considered our stacking sample against an extrapolation of the $z \sim 2.3$ $M_\ast-\rm{SFR}$ SFMS of \citet{Sanders2021}, finding the sample to scatter on either side of this relation. Our stacking sample has a median SFR (and SFR interquartile range) of $\rm{SFR}^{\rm{med}}_{\rm{H}\alpha} = 2.25^{+2.15}_{-1.26}\ \rm{M_\odot\, yr^{-1}}$, which lies $\Delta\log(\rm{SFR}) \approx 0.19$ dex above this SFMS at fixed $M_\ast$ ($\log(M_\ast/\rm{M_\odot})_{\rm{med}} = 8.29^{+0.51}_{-0.43}$), corresponding to a bias in O/H of $\Delta\log(\rm{O/H}) \approx -0.06$ dex via the FMR, well within even our statistical metallicity uncertainty ($\sigma_{\rm{stat}} \sim 0.12$ dex). We conclude that our stacking sample is not largely biased in SFR or O/H and is thus, on average, representative of typical, star-forming, $z \sim 2.3$ dwarf galaxies with stellar masses between $10^8 \lesssim M_\ast/\rm{M_\odot} \lesssim 10^9$. Our sample serves as an initial extension of representative, statistical, spectroscopic surveys at $z \sim 2.3$ into the dwarf galaxy mass regime.
    
    \item We analyzed the applicability at $z \sim 2.3$ of several locally-calibrated, oxygen-based, strong-line metallicity relations from the literature. We find that at $12+\log(\rm{O/H})_{\rm{direct}} = 7.88^{+0.25}_{-0.22}$, our stack lies in the metallicity-insensitive ``turnover" region of the $\rm{O_3}$ and $\rm{R_{23}}$ calibrations, signalling their ineffectiveness for metallicity estimation of typical, $z \sim 2.3$, dwarf galaxies. When considering the $\rm{O_{32}}$ and $\rm{O_2}$ indices together, our stack's metallicity is most accurately reproduced (within $\lesssim 0.12$ dex) at fixed strong-line ratio by the local reference calibrations of \citet{Bian2018}, in agreement with that seen for the $z \sim 2.4$ KBSS-LM1 composite of \citet{Steidel2016}, who also selected their sample independent of line-strength. We generally disagree with the conclusions of \citet{Sanders2020}, who argue that their $z \sim 2.2$ auroral-line sample favors the high-$z$ analog calibrations of \citet{Bian2018}. While both samples are at similar redshift and metallicity, we argue that our discrepancy in conclusion is due to sample selection effects as well as biases in the low-metallicity strong-line calibration samples. Indeed, by being selected for having a detection of a $T_e$-sensitive auroral-line, the sample of \citet{Sanders2020} is strongly biased in SFR, $\rm{O_{32}}$, and $\rm{EW_0}$([\ion{O}{3}] $\lambda$5007) relative to typical, $z \sim 2.3$, star-forming galaxies. 
    
    \item At the median stellar mass of our stacking sample, log($M_\ast$/$\rm{M_\odot}$) = $8.29^{+0.51}_{-0.43}$, we compared our composite direct metallicity, $12+\log(\rm{O/H}) = 7.88^{+0.25}_{-0.22}$, against the $z \sim 2.2$ direct-method MZR of \citet{Sanders2020}, the $z \sim 2.3$ strong-line MZR of \cite{Sanders2021}, and the $z \sim 2.3$ photoionization-model-based MZR of \citet{Strom2022}. After correcting for the slight SFR bias of the stacking sample, we find that our $z \sim 2.3$ stack lies $\sim 0.07$ dex above the direct-method MZR and $\sim 0.06$ dex below the strong-line MZR at fixed $M_\ast$, well within our uncertainties. Our stack lies $\approx 1\sigma$ below the Strom et al. MZR. In constraining the slope of the MZR, we defer to the direct-method MZR, with which we show excellent agreement, as metallicities for this relation and our composite were calculated consistently. Therefore, we suggest that the slope of the $z \sim 2.3$ MZR is that given by this direct-method relation of \citet{Sanders2020}, $\beta=0.37$.
    
    \item We also compared our composite, and the MZRs derived from observations, to the $z=2.30$ MZR from the FIRE simulations \citep{Ma2016} as well as to the $z=2$ and $z=3$ MZRs from the IllustrisTNG100 simulations \citep{Torrey2019}. At the stack's fiducial median stellar mass, log($M_\ast$/$\rm{M_\odot}$) = $8.29^{+0.51}_{-0.43}$, our composite is consistent within uncertainties with both sets of simulations. However, when recalculating our sample stellar masses assuming more realistic non-parametric SFHs, the median stellar mass is increased to log($M_\ast$/$\rm{M_\odot}$) = $8.92^{+0.31}_{-0.22}$, moving the stack into excellent agreement with the FIRE MZR and $\sim 1.5\sigma$ below the $2 \leqslant z \leqslant 3$ IllustrisTNG MZR. This tension with IllustrisTNG is in part caused by a ``bump" in its MZR deriving from a constant minimum wind velocity ($v_{\rm{min}}=350\ \rm{km}\ \rm{s^{-1}}$) applied to galaxies with $M_\ast \lesssim 10^9\ \rm{M_\odot}$. Between our stack with recalculated masses and the direct-method MZR of \citet{Sanders2020}, we suggest that the low-mass end of the MZR does not contain this bump.
    
    \item Our $z \sim 2.3$ dwarf galaxy composite was compared to the locally-defined, direct-method FMR in order to test the relation's redshift-invariance. We made this comparison via the FMR projection proposed by \citet{Mannucci2010} and given in Equation \ref{equ:fmr_projection}, with $\alpha=0.63$ \citep{Sanders2017}. At $\mu_{0.63} = 8.07$, calculated with the stacking sample's median stellar mass, log($M_\ast$/$\rm{M_\odot}$) = 8.29, and median SFR, log(SFR) = 0.353, we find our composite to lie $\sim 0.4\sigma$ or $\sim 0.10$ dex below the \citet{Sanders2017} best-fit linear relation in direct-method O/H$\,-\,\mu_{0.63}$ space; this relation is fit to the $z \sim 0$ DIG-corrected stacks of \citet{Andrews&Martini2013}. We therefore agree with many in the literature who suggest that the FMR is redshift-invariant within $\sim 0.1$ dex at fixed $M_\ast$ and SFR from $z \sim 0-2.3$.
\end{enumerate}

This study compliments other larger spectroscopic surveys of representative, star-forming galaxies at $z \sim 2$ and $M_\ast \gtrsim 10^9\ \rm{M_\odot}$ by serving as an initial extension into the dwarf galaxy mass regime. In analyzing our sample of representative dwarf galaxies, we are able to use a direct oxygen abundance to provide initial constraints on the low-mass slope of the high-$z$ MZR and probe scarcely-studied parameter space of the FMR. While our sample size is small, our work provides a reference point for future statistical studies of high-$z$ dwarf galaxies with the newly-operational \textit{James Webb Space Telescope}, which will greatly increase our understanding of the processes responsible for galaxy formation and evolution.

\begin{acknowledgments}

This material is based upon work supported by the National Science Foundation under Grant No. 1617013.

The authors wish to recognize and acknowledge the very significant cultural role and reverence that the summit of Maunakea has always had within the indigenous Hawaiian community.  We are most fortunate to have the opportunity to conduct observations from this mountain. 

\end{acknowledgments}

\facilities{Keck:I (MOSFIRE), \textit{HST} (WFC3, ACS)}

\appendix

\section{Consistent Fits of the Strong-Line Metallicity Calibrations of Bian+18} \label{sec:b18_appendix}
We note that while \citet{Bian2018} provide fits to their high-$z$ analog stacks for the strong-line indices of $\rm{R_{23}}$, $\rm{O_{32}}$, and $\rm{O_3}$ (with [\ion{O}{3}] $\lambda$4959 added to [\ion{O}{3}] $\lambda$5007 in the latter two indices unlike in this work; see \citet{Bian2018} and \citet{Sanders2021}, Footnote 17), they do not provide fits for the $\rm{O_2}$ index or to the local reference stacks. Therefore, we used orthogonal distance regression and the stated flux and $T_e$-based metallicity values given in \citet{Bian2018} to consistently fit both samples for each line index considered in Figure \ref{fig:slr_dir_metal} and Section \ref{subsec:SLR_O/H}. As in \citet{Bian2018}, third-order polynomials were assumed for the $\rm{R_{23}}$ and $\rm{O_3}$ high-$z$ analog fits; we adopted the same functional form for the $\rm{O_2}$ high-$z$ analog fit as well. For the fits of the local reference stacks and these same strong-line indices, we instead assumed second-order polynomial functions due to there only being 4 bins with estimated direct metallicities compared to 6 bins for the high-$z$ analog sample. For the $\rm{O_{32}}$ index, as in \citet{Bian2018}, we fit a linear functional form to both the local reference and high-$z$ analog stacks. The coefficients of our fits are given in Table \ref{tab:bian_coeffs} below. 
\vspace{0.5cm}

\begin{deluxetable}{lrrrr}[ht!]
\vspace{0.1cm}
\tablecaption{Coefficients of the Refit Bian+18 Relations \label{tab:bian_coeffs}}
\tablecolumns{5}
\tablenum{4}
\tablewidth{\textwidth}
\renewcommand{\arraystretch}{1.2}
\setlength{\tabcolsep}{6.8pt}
\tablehead{
\colhead{Ratio} &
\colhead{$c_0$} &
\colhead{$c_1$} &
\colhead{$c_2$} &
\colhead{$c_3$}
}
\startdata
\hline
\multicolumn{5}{c}{Local Reference Relations} \\
\hline
$\rm{O_{32}}$ & 11.1767  & -1.3414 &         &  \\
$\rm{O_3}$    & -57.7632 & 14.9708 & -0.9578 &  \\
$\rm{O_2}$    & -25.2699 & 5.3889  & -0.2750 &  \\
$\rm{R_{23}}$ & -26.4726 & 6.9012  & -0.4345 &  \\
\hline
\multicolumn{5}{c}{High-$z$ Analog Relations} \\
\hline
$\rm{O_{32}}$ & 14.5895  & -1.7287   &         &         \\
$\rm{O_3}$    & 117.8668 & -48.8336  & 6.7607  & -0.3107 \\
$\rm{O_2}$    & -80.9802 & 14.1526   & 0.0755  & -0.0723 \\
$\rm{R_{23}}$ & 263.2135 & -101.2448 & 13.0313 & -0.5591 
\enddata
\tablecomments{Ratio definitions given in Section \ref{subsec:SLR_O/H}.}
\end{deluxetable}

\bibliography{stack_paper_v2.bib}{}
\bibliographystyle{aasjournal.bst}

\end{document}